\documentclass[twocolumn]{aastex631}
\usepackage{mathptmx}

\shorttitle{Perpendicular planets}
\shortauthors{Albrecht et al.}

\begin{document}

\title{A Preponderance of Perpendicular Planets}

\author[0000-0003-1762-8235]{Simon H.\ Albrecht}
\affil{Stellar Astrophysics Centre, Department of Physics and Astronomy, Aarhus University, Ny Munkegade 120, 8000 Aarhus C, Denmark}

\author[0000-0003-2173-0689]{Marcus L.\ Marcussen}
\affil{Stellar Astrophysics Centre, Department of Physics and Astronomy, Aarhus University, Ny Munkegade 120, 8000 Aarhus C, Denmark}

\author[0000-0002-4265-047X]{Joshua N.\ Winn}
\affiliation{Department of Astrophysical Sciences, Peyton Hall, 4 Ivy Lane, Princeton, NJ 08544, USA}

\author[0000-0001-9677-1296]{Rebekah I.\ Dawson}
\affiliation{Department of Astronomy \& Astrophysics, Center for Exoplanets and Habitable
Worlds,The Pennsylvania State University, University Park, PA 16802, USA}

\author[0000-0001-7880-594X]{Emil Knudstrup}
\affil{Stellar Astrophysics Centre, Department of Physics and Astronomy, Aarhus University, Ny Munkegade 120, 8000 Aarhus C, Denmark}

\begin{abstract}
Observing the Rossiter-McLaughlin effect during a planetary transit allows the determination of the angle $\lambda$ between the sky projections of the star's spin axis and the planet's orbital axis.  Such observations have revealed a large population of well-aligned systems and a smaller population of misaligned systems, with values of $\lambda$ ranging up to 180$^\circ$. For a subset of 57 systems, we can now go beyond the sky projection and determine the 3-d obliquity $\psi$ by combining the Rossiter-McLaughlin data with constraints on the line-of-sight inclination of the spin axis. Here we show that the misaligned systems do not span the full range of obliquities; they show a preference for nearly-perpendicular orbits ($\psi=80-125^\circ$) that seems unlikely to be a statistical fluke. If confirmed by further observations, this pile-up of polar orbits is a clue about the unknown processes of obliquity excitation and evolution.
\end{abstract}

 
\section{Introduction} \label{sec:intro}

An interesting development in exoplanet science was the discovery
that a star's direction of rotation need not be aligned with the orbital
motion of its planets \citep{hebrard2008,Winn+2009_X03,queloz2010}.
This fact came to light through observations of the
Rossiter-McLaughlin (RM) effect, a spectroscopic anomaly during planetary
transits that depends on the degree of spin-orbit
alignment \citep{triaud2017}. 

\begin{figure}
  \begin{center}
    \includegraphics[width=1\columnwidth]{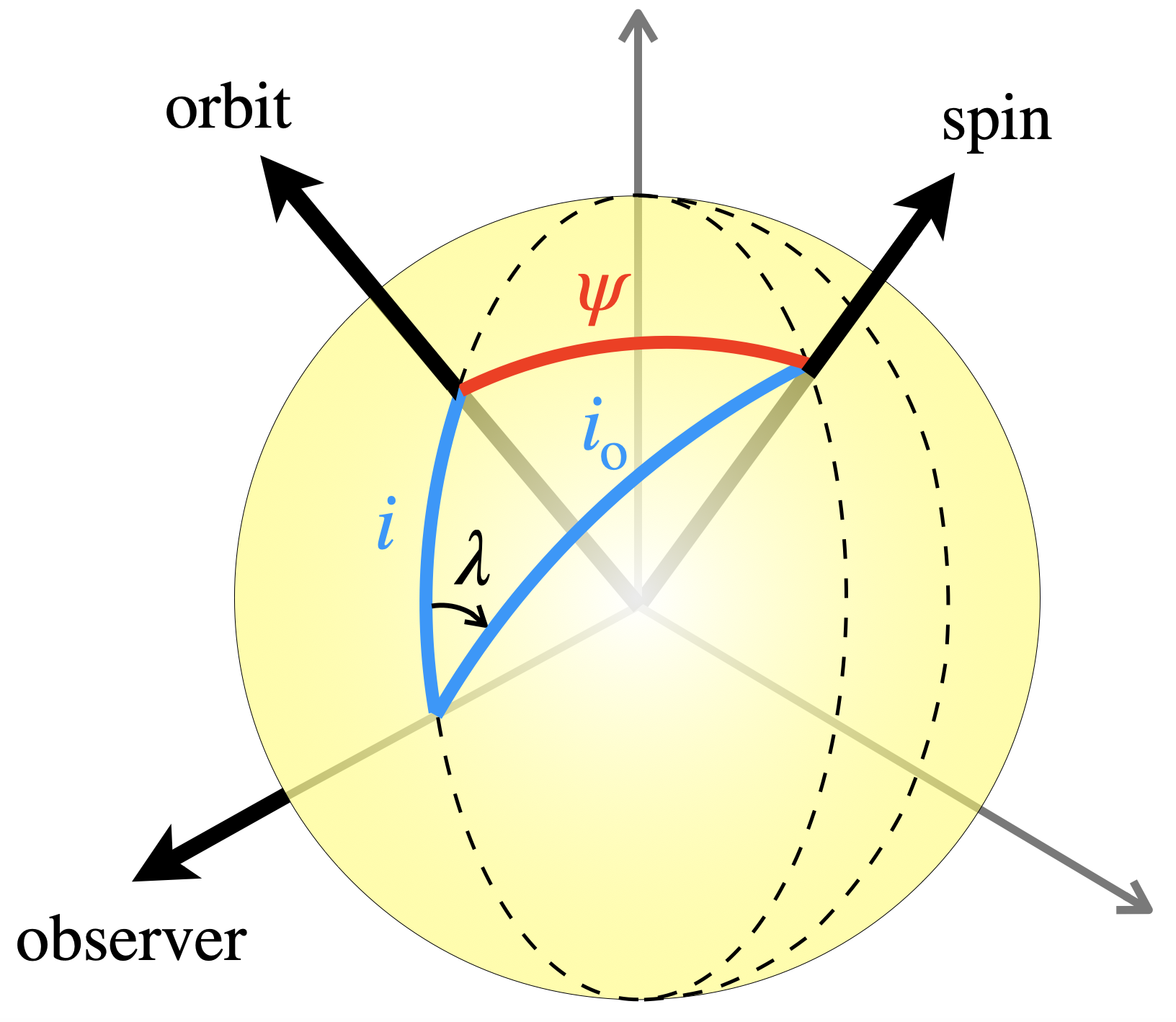}
    \caption{
    \label{fig:geometry} 
    \textbf{Geometry of the problem.}
    Shown are the obliquity $\psi$,
    the sky-projected obliquity $\lambda$, the star's inclination $i$,
    and the orbital inclination $i_\mathrm{ o}$
    (which is always near 90$^\circ$ for a transiting planet).
    Modeled after a similar figure by \cite{perryman2011}.
    }
  \end{center}
\end{figure}

An obstacle to the interpretation of these results is that
the RM effect depends mainly on the angle $\lambda$
between the {\it sky projections} of the rotational
and orbital axes, but the physically important angle is the obliquity $\psi$ measured between the axes in three dimensions.
See Figure~\ref{fig:geometry} for a diagram of the angles, and
\cite{fabrycky_winn2009} for more details on the geometry.

\begin{figure}
  \begin{center}
    \includegraphics[width=1\columnwidth]{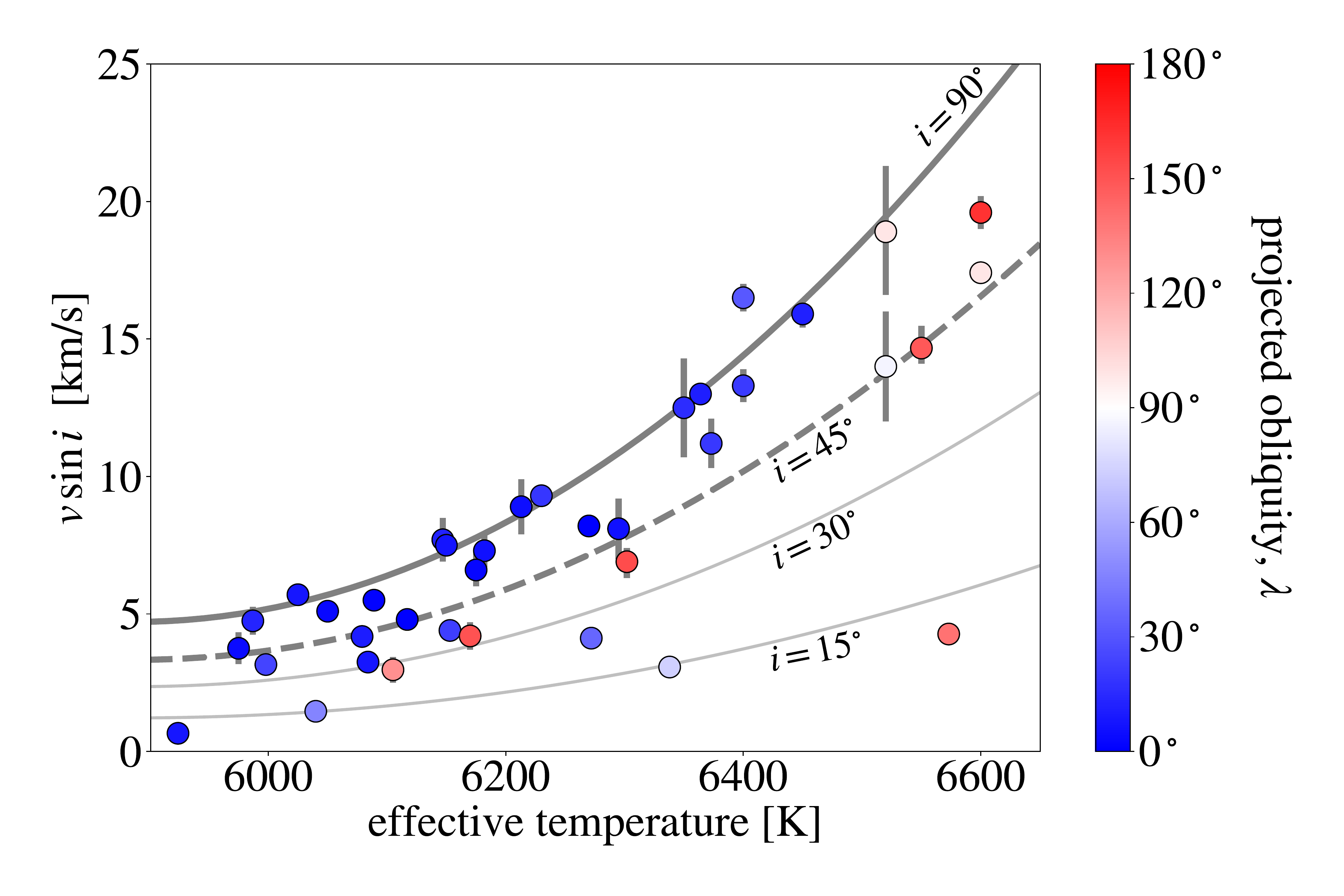}
    \caption{
    \label{fig:vsini_teff} 
    \textbf{Suspiciously low \emph{v}\,sin\,\emph{i}.}
    Shown are all stars for which (i) $\lambda$ has been
    measured, (ii) the effective temperature is within
    the plotted range, and (iii) the parameter $(m/M)(R/a)^3$
    is less than 0.002, to avoid tidally spun-up stars.
    Whenever $\lambda>90^\circ$ (white and red points),
    the $v\sin i$ is abnormally low.
    The thick line is the fitting function $v(T_\mathrm{ eff})$
    from \cite{Louden+2021}, and the other lines are scaled
    for various inclinations.
    Of the 8 stars with $\lambda>90^\circ$, five (63\%) have inclinations $<$\,$45^\circ$, more than the 30\% we would expect from
    a randomly-oriented population.
    }
  \end{center}
\end{figure}

To determine $\psi$, observations of the RM effect must be
supplemented with information about the inclination $i$ of the star's
rotation axis with respect to the line of sight.
For example, a measurement of
the star's projected rotation velocity $v\sin i$ can
be combined with the expected value
of the rotation velocity $v$ based on the star's mass and age
to arrive at a constraint on $\sin i$ \citep{schlaufman2010}.
Recently, \cite{Louden+2021} helped to set expectations
for the rotation velocities of stars with effective
temperatures between 5900 and 6600\,K by measuring
the $v\sin i$ distribution of a randomly-oriented sample of stars.
In reviewing the RM data for such stars,
we noticed that whenever $\lambda$ exceeds $90^\circ$,
the inclination tends to be quite low, indicating
a nearly polar orbit as opposed to a retrograde orbit.
This pattern, shown in Figure~\ref{fig:vsini_teff}, made
us wonder if the 3-d obliquity distribution
shows a concentration near $90^\circ$ even though
the projected obliquity distribution spans
the full range from 0$^\circ$ to 180$^\circ$.

There is a growing number of stars for which the
inclination can be determined directly from the data,
even without prior expectations for the rotation velocity.
In this Letter, we collect and analyze all these systems (Section~\ref{sec:sample}),
and show that the distribution of $\psi$
does indeed have a peak
near 90$^\circ$ (Section~\ref{sec:distribution}).
We perform tests
of the statistical significance of
this pattern (Section~\ref{sec:tests}) and consider
possible biases (Section~\ref{sec:bias}).
Finally, we speculate on possible physical explanations (Section~\ref{sec:discussion}).

\section{Sample selection}
\label{sec:sample}

\begin{figure*}
\begin{center}
\includegraphics[width=1.0\textwidth]{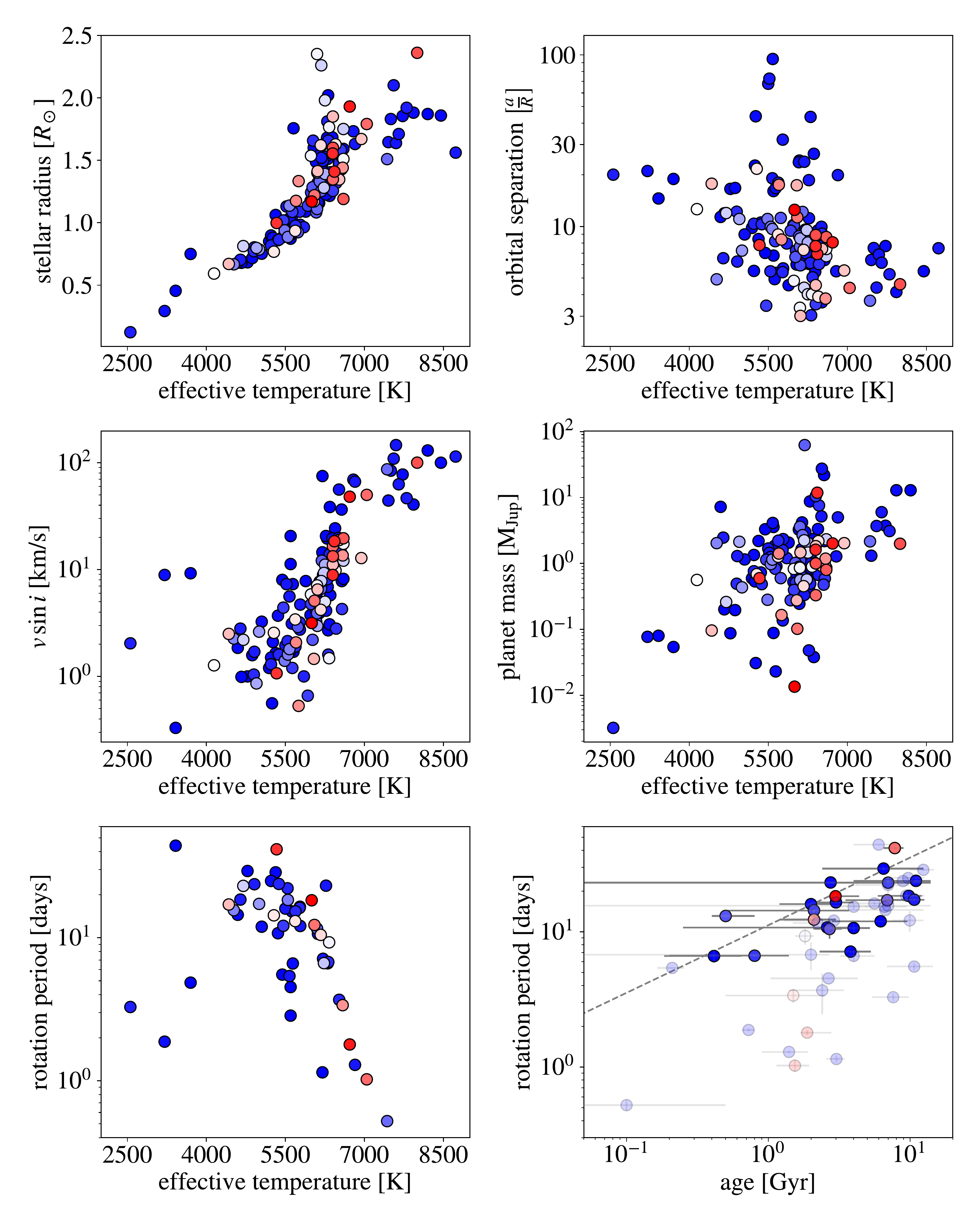}
\end{center}
\vskip -0.2in
\caption{{\bf Stars with measurements of sky-projected obliquity.}
The color scale indicates $\lambda$, from blue for $0^\circ$ to red for $180^\circ$.
As a function of the star's effective temperature,
the left column shows the star's radius (top),
sky-projected rotation velocity (middle),
and rotation period when available (bottom).
The right column shows the planet's orbital separation
in units of the stellar radius
(top) and mass (middle). The lower right panel shows
the star's rotation period vs.\ age,
with a line
indicating the Skumanich law $P\propto \sqrt{t}$.
The darker data points for which the stars are expected to
obey the Skumanich law: $T_{\rm eff} = 4500$--6000\,K, $(m/M)(R/a)^3<0.002$.
\label{fig:sample}
}
\end{figure*}

Our starting point was the online
database \href{https://www.astro.keele.ac.uk/jkt/tepcat/}{TEPCAT}
\citep{southworth2011}, which includes a compilation of results
from RM observations.  
To the 155 systems that were in this
database on 5~Jan~2021, we added K2-290 \citep{Hjorth+2021}
for a total of 156.

Our main method for determining the stellar inclination angle
was to combine
measurements of the star's projected rotation velocity ($v\sin i$),
radius ($R$), and rotation period ($P_\mathrm{ rot}$):
\begin{equation}
    \label{eq:vsini_over_v}
    \sin i = \frac{v\sin i}{v} = \frac{P_{\rm rot}\, v\sin i}{2\pi R}.
\end{equation}
The limiting factor was knowledge of the rotation period.
A literature search turned up 38 cases in which the rotation period
had been determined through photometric monitoring.
We also analyzed all the available light curves from the
{\it Kepler}, {\it K2}, and {\it TESS} missions using the
autocorrelation method \citep{McQuillanMazehAigrain2013},
which led to rotation period measurements for 28 stars.
Of these, 15 were in agreement with values reported in the literature, and 13 had not been reported previously.
All together we had 51 stars with measured rotation periods.

In addition, there are a few stars for which the
stellar inclination angle was determined using a different method.
For HAT-P-7, we used the result from the asteroseismic method, based on the relative amplitudes
of the members of rotationally-split multiplets in the oscillation spectrum.
For KELT-9, KELT-17, Kepler-13, MASCARA-4, and WASP-189, we used results from
the gravity-darkening method, based on modeling the distortions to the
transit light curve due to the star's equator-to-pole intensity gradient.
Table~\ref{tab:systems} gives all the data and citations to the literature.

This made for a total of
57 systems for which we could determine the 3-d obliquity. 
Figure~\ref{fig:sample} displays the key properties of the sample.
Throughout this figure (and the two figures to follow),
the color of each data point
conveys $\lambda$, with
blue for $0^\circ$, white for $90^\circ$, and red for $180^\circ$.
The upper left panel shows that the stars range from M dwarfs to A stars.
The middle left plot shows $v\sin i$ vs.\ effective temperature,
including the sharp rise around 6250\,K (the ``Kraft break'').
The lower left panel shows the subset of stars with 
measured rotation periods.
The lower right panel shows rotation period as a function of main-sequence age,
as estimated from stellar-evolutionary models. For reference,
we highlighted the stars we expect to conform
to the Skumanich law, $P_{\rm rot}\propto \sqrt{t}$, indicated
by the dashed line. Specifically,
the highlighted stars have effective temperatures between 4000 and 6500\,K
and are unlikely to have been tidally spun up because
$(m/M)(R/a)^3 > 0.002$ (corresponding to $a/R > 8$ for
a Jupiter-mass planet around a Sun-like star).

\section{Obliquity distribution}
\label{sec:distribution}

For the stars with measured rotation periods, we used the
method of \cite{MasudaWinn2020} to determine the posterior probability
distribution for the obliquity, using the equation
\begin{equation}
\label{eq:cospsi}
    \cos\psi = \sin i\,\sin i_{\rm o}\,\cos\lambda + \cos i\,\cos i_{\rm o}.
\end{equation}
There are two solutions because the available observations cannot distinguish $i$ from $180^\circ-i$, nor can they distinguish ($i_{\rm o}$, $\lambda$) from $(180^\circ - i_{\rm o}$, $-\lambda$). The two solutions are closely spaced because $i_{\rm o} \approx 90^\circ$.

In a few cases, the values of $\lambda$ reported in the literature had uncertainties
less than a degree. Out of concern about systematic errors, we imposed a minimum
uncertainty of $1^\circ$ in our analysis. For the same reason, we imposed a minimum uncertainty
of 0.1\,km/s in $v\sin i$.
In all cases for which $\psi$ had
been reported previously in the literature, our results were in agreement.
In four of these cases --- HAT-P-11, K2-290, Kepler-63, and WASP-107 ---
we adopted the previously reported value because
it was based on more data.

\begin{figure*}
\begin{center}
\includegraphics[width=1.0\textwidth]{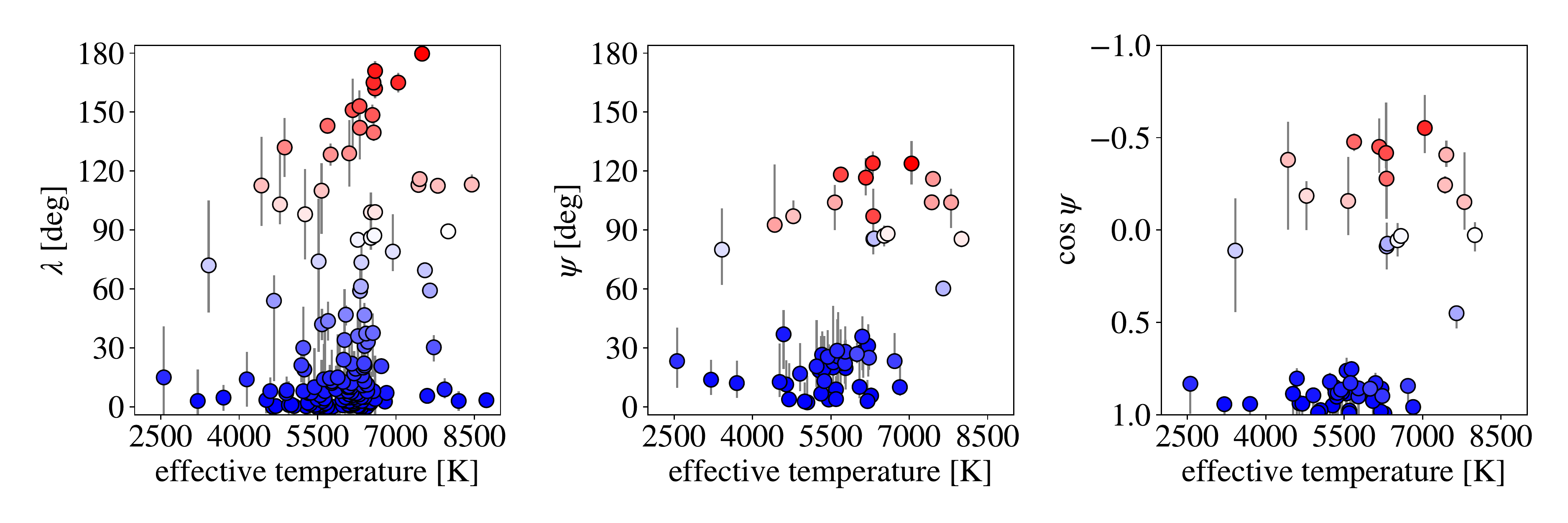}
\end{center}
\vskip -0.2in
\caption{{\bf Obliquity distribution.} Shown as a function of the star's
effective temperature are the sky-projected obliquity (left), the 3-d obliquity (middle),
and the cosine of the 3-d obliquity (right). \label{fig:psi}}
\end{figure*}

\begin{figure}
\begin{center}
\includegraphics[width=1.0\columnwidth]{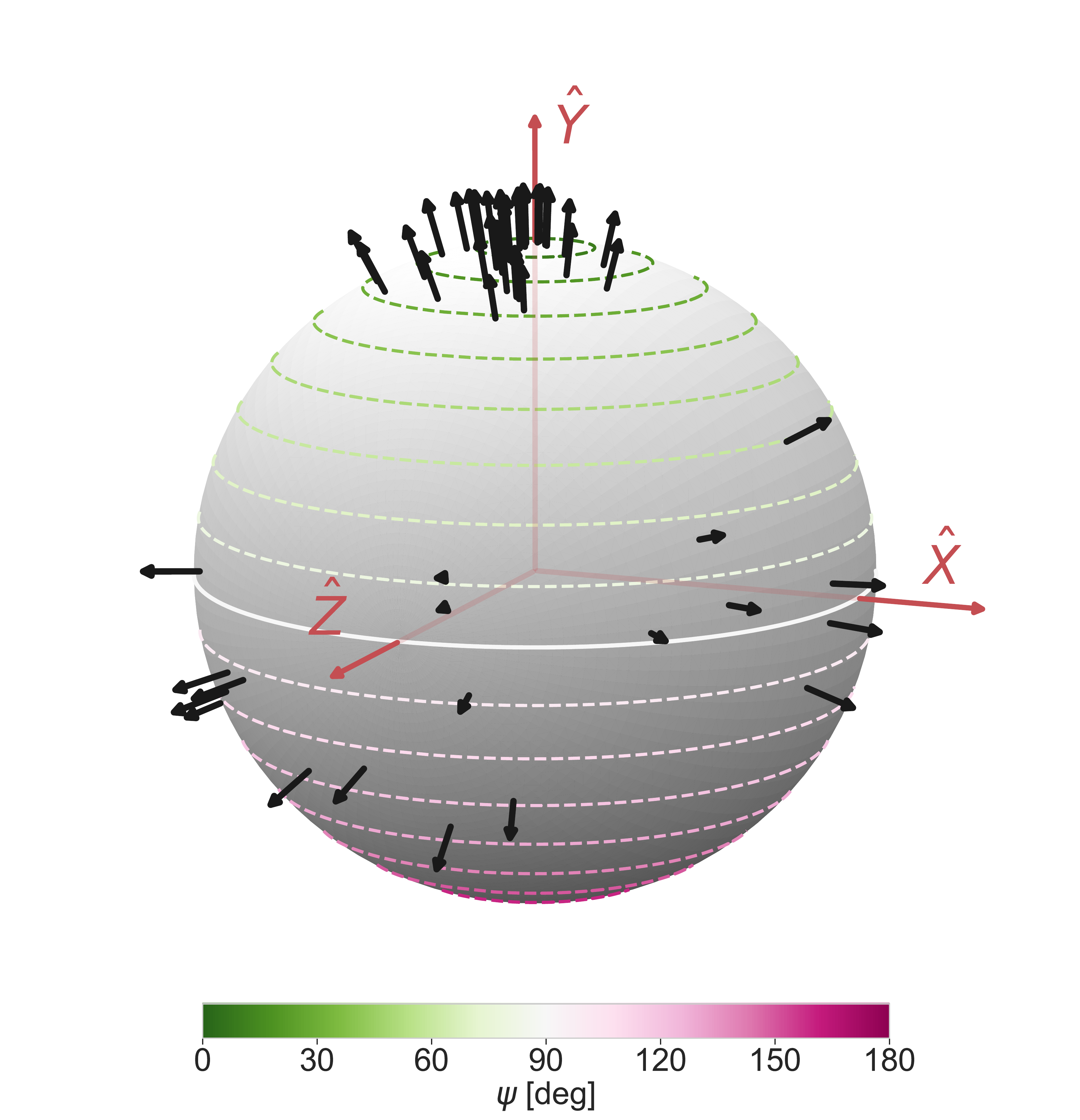}
\end{center}
\caption{\textbf{Obliquity distribution in 3-d.}
The $z$-axis is the line of sight, and the $y$-axis is the orbital axis.
The arrows represent the orientations of
the stellar spin axes in our sample, and the colors of the latitudinal lines convey the obliquity. For plotting purposes, we broke the two-way degeneracy by choosing $i$ and $i_{\rm o}$ to be $\leq$\,$90^\circ$
and taking the signed value of $\lambda$ from the literature.
\label{fig:3d}}
\end{figure}

Figure~\ref{fig:psi} shows the results as function of effective temperature.
For ease of visual interpretation, instead of showing both solutions for $\psi$,
we show the results assuming $i_{\rm o}=90^\circ$.\footnote{The median spacing between
the two degenerate solutions is $0.2\,\sigma$ where $\sigma$ is the statistical uncertainty,
and the maximum spacing is $2\,\sigma$.}
The left panel of Figure~\ref{fig:psi} shows the
distribution of $\lambda$, which ranges from $0^\circ$ to $180^\circ$.
The middle panel shows the distribution of $\psi$, in which there are two groups:
38 well-aligned systems with $\psi\lesssim 35^\circ$ 
and 18 systems with $\psi$ in between 80$^\circ$ and 125$^\circ$.
The right panel shows the distribution of $\cos\psi$, which is
easier to interpret visually 
because randomly oriented stars would have a uniform
distribution in $\cos\psi$.
As a less exact and more visual representation of the results,
Figure~\ref{fig:3d} shows a 3-d representation of the stellar
spin orientations. 

\begin{figure*}
\begin{center}
\includegraphics[width=1.0\textwidth]{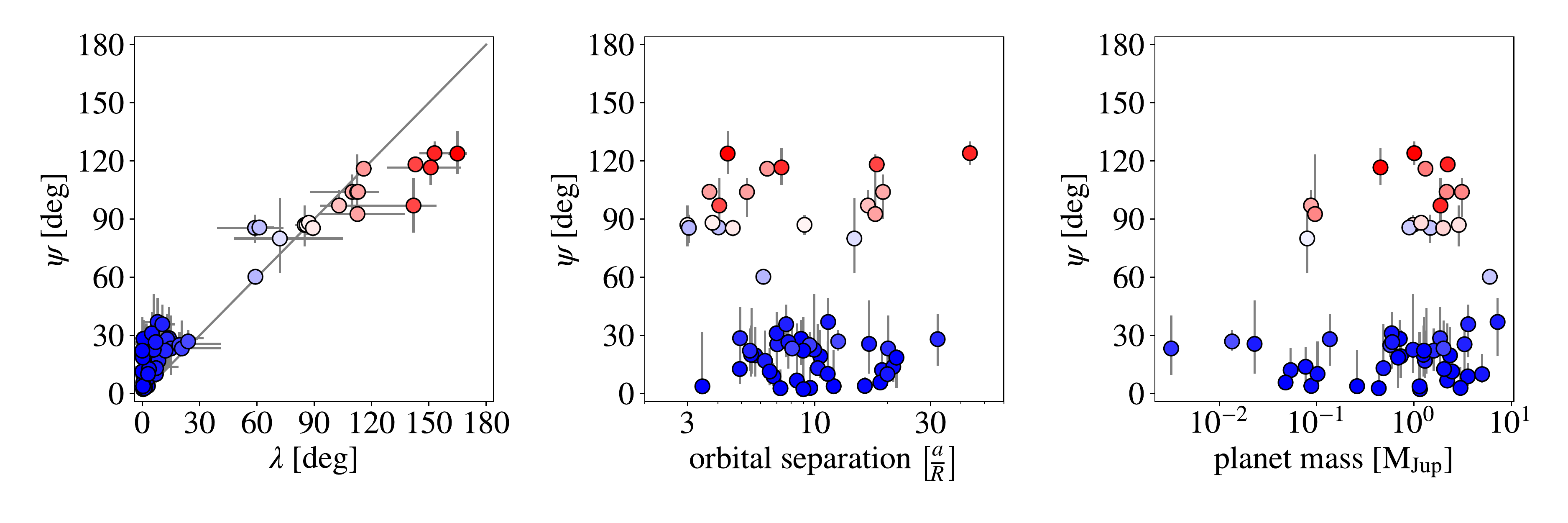}
\end{center}
\vskip -0.2in
\caption{{\bf No trends seen.} The 3-d obliquity is plotted as a function
of the sky-projected obliquity (left), the planet's orbital separation (middle),
and the planet's mass (right). \label{fig:psi2}}
\end{figure*}

Seeking clues to the origin of this pattern,
we tried to find something else the misaligned stars all have in common.
Nothing stood out (see Figure~\ref{fig:psi2}).
The misaligned group includes stars of spectral types A through M,
orbital distances from 3 to 40 stellar radii,
and planet masses from 0.1 to 3 M$_{\rm Jup}$.
The three Neptune-mass planets GJ\,436, HAT-P-11, and 
WASP-107 all have perpendicular orbits,
as do the six hot Jupiters
HAT-P-7, KELT-9, and WASP-7/76/189.
The subgroup of systems with $\psi\approx 110^\circ$ consists
of hot Jupiters around A, F, and G stars as well as
the two-transiting-planet system K2-290.
The lone star in between the aligned and misaligned
groups is Kepler-13, with $\psi = 60^\circ$.
This star also has the most massive planet 
among the misaligned stars (4.9--8.1\,M$_{\rm Jup}$,~\citealt{Shporer+2014}).
This could be a coincidence, although it does
conform to the previously noted pattern that the stars with
the most massive planets are rarely
found to have $\lambda>90^\circ$ \citep{hebrard2011b}.

\section{Statistical tests}
\label{sec:tests}

We performed several statistical tests of the
``null hypothesis'' that the misaligned stars are randomly oriented
instead of being clustered around $\psi\approx 90^\circ$.
Defining misaligned stars as having $\cos\psi<0.75$
($\psi>41^\circ$), the null hypothesis would lead to the expectation that
$\cos\psi$ is uniformly distributed between
$-1$ and $0.75$.

One way to characterize the distribution of the misaligned stars
is by the observed dispersion of $\cos\psi$ around zero:
\begin{equation}
\label{eq:dispersion}
    \sigma_0 \equiv \sqrt{ \frac{1}{N} \Sigma_{i=1}^{N} \left(\cos\psi_i \right)^2},
\end{equation}
where $N=19$. If $\cos\psi$ were distributed uniformly for the misaligned
stars, how often would $\sigma_0$ be at least as small as is observed?
To construct the expected distribution of $\sigma_0$ under this null hypothesis,
we repeatedly drew 19 numbers from a uniform
random distribution between $-1$ and $0.75$ and calculated
$\sigma_0$ in each case.
Then, in each of $10^5$ trials, we constructed a realization of the data
by (i) choosing one of the two solutions of Equation~\ref{eq:cospsi} at random
for each data point, (ii) drawing a value of $\cos\psi$ from the observational
posterior. For each realization of the data, we calculated
the $p$-value of the null hypothesis. The median $p$-value was $3.1\times 10^{-3}$.

We obtained similar results when replacing $\sigma_0$ by the standard deviation, i.e., when
testing for clustering of $\cos\psi$ around the mean value, rather than around zero.
Through similar Monte Carlo simulations, we found that the median $p$-value in that case is
$9.6\times 10^{-4}$.

The left column of Table 2 summarizes the results of these tests.
While the $p$-values are low enough to
reject the null hypothesis according to customary criteria,
we acknowledge that the choices of the
specific tests and the threshold value of $\cos\psi$ to qualify as
``misaligned'' were devised {\it after} seeing the data.
As always in such cases, caution is warranted,
and there is no substitute for getting additional data.

\begin{deluxetable}{lcc}
\label{tbl:pvalues}
\tablecaption{$p$-values.}
\tablehead{
  Test & including  & excluding \\
       & hottest stars & hottest stars
}
\startdata
disp.\ around $90^\circ$  & $0.00310$ &  $0.0100$ \\
disp.\ around mean        & $0.00096$ &  $0.0044$ \\ 
\enddata
\end{deluxetable}

\section{Possible biases}
\label{sec:bias}

The basic result of our study is that $|\!\cos\psi|$ tends to be small
for the misaligned systems. 
Suppose that systematic errors in our input data have caused
the inferred value of $|\!\cos\psi|$ to be biased
by a factor $f$, i.e., when we infer $\cos\psi = x$ the true value is $x(1+f)$.
How large would $f$ need to be for the evidence for clustering around $90^\circ$
to go away?  We answered this question through another Monte Carlo simulation:
we searched for the
value of $f$ that causes the median $p$-value
to rise to 0.05 in the standard-deviation test
described in the previous section.
The answer is $f\approx 0.30$.

It seems unlikely that the inputs are biased at the 30\% level.
For the rotation period method, the obliquity is determined via the equation
\begin{equation}
    \cos\psi \approx \cos\lambda\,\frac{P_\mathrm{ rot}\,v\sin i}{2\pi R}.
\end{equation}
Therefore, a bias toward low $|\!\cos\psi|$ could be caused by:
\begin{enumerate}
    \item Underestimating $v\sin i$. We adopted the results from
    analyses of RM data. Cross-checks with results based
    on spectral-line broadening indicate agreement within 10\%
    with no discernible bias.
    \item Overestimating $R$.  The evolutionary models used to
    determine $R$ are not perfect, but for these well-characterized
    stars any bias is likely to be $\lesssim 10\%$.
    \item Underestimating $P_\mathrm{ rot}$. 
    Differential rotation probably introduces biases at the $\lesssim 10\%$ level.
    For example, the photometric period might be associated with
    features at different latitudes than those that contribute to
    the determination of $v\sin i$.
\end{enumerate}

Another possible problem is that
sometimes a photometric period is not the rotation period, but rather
one-half of the rotation period, because spots
occur on opposite hemispheres. Supposing this factor-of-two error occurred
for $n$ stars out of the sample of 19, we found through Monte Carlo
simulation that we would need $n\gtrsim 6$ to cause the $p$-value of the standard-deviation test to rise to 0.05.
We cannot exclude this possibility, though it
seems doubtful given the good agreement between the periods
we derived and the periods reported in the literature based on
earlier data. Also, in many cases the measured rotation
period conforms with expectations given the star's mass
and age (lower panels of Figure~\ref{fig:sample}).

We also tried omitting the stars with $T_\mathrm{ eff}>7000$\,K
from our statistical tests, for two reasons.  First, for such hot stars,
the photometric period might actually be a pulsation period
instead of the rotation period. Second, our sample might be biased
against hot stars with $\psi=0^\circ$ or $180^\circ$
because such stars would have larger $v\sin i$ values,
making it more difficult to perform Doppler spectroscopy
and confirm a planetary signal.
The resulting $p$ values, given in the second column of Table~\ref{tbl:pvalues},
are $\lesssim$\,$0.01$.

\section{Discussion}
\label{sec:discussion}

The literature contains a previous observational hint
that nearly-polar orbits are common,
as well as several theoretical scenarios that might be relevant.
On the observational side,
\citet{mazeh2015} found statistical evidence that
stars with effective temperatures between about 6000 and 6500\,K
have high obliquities, with a possible preference for polar
orbits.
This tentative conclusion was based on the observation
that the amplitude of photometric variations associated with rotation
was lower for stars with transiting planets than for randomly-oriented stars.
The correspondence with our sample is not exact, though:
our stars span a wider range of effective temperatures,
and our planets are generally larger than
those analyzed by \citet{mazeh2015}.

On the theoretical side, four scenarios that could lead to obliquities
near 90$^\circ$ are:
\begin{enumerate}

\item \textbf{Tidal dissipation} is usually thought to damp obliquities
to $0^\circ$ but in some cases can cause the obliquity to linger at $90^\circ$.
\cite{lai2012} showed this can happen when the damping is
dominated by the dissipation of inertial waves driven
in the convective zone by Coriolis forces \citep[see also][]{rogers2013b,anderson2021}.
This theory might account for a few near-perpendicular systems.
However, many systems have properties that seem incompatible with this theory:
six have stars that lack convective zones ($T_{\rm eff} > 7000$ K) and seven have
orbital separations beyond 10 stellar
radii where tides are expected to be negligible. Furthermore, 12 systems do not satisfy
the criterion $P_{\rm rot} > 2P_{\rm orb}$ necessary to prevent tidal orbital decay,
and 5 have large angular momentum ratios $L_{\rm orb}/L_{\rm spin}$ which would
cause the obliquity to stall at $180^\circ$ instead of $90^\circ$.
 
\item \textbf{Von Zeipel-Kozai-Lidov cycles}, often invoked to explain hot Jupiters as the outcome
of high-eccentricity tidal migration, were predicted to lead an obliquity distribution with
a peak near $115^\circ$ \citep{fabrycky2007}, a good match to the data.
The predicted distribution also has a peak near $35^\circ$,
while the low-obliquity group in our sample has $\langle\psi\rangle = 15^\circ$.
Subsequent studies have shown that the predicted obliquity distribution 
depends on the orbital parameters of the perturber \citep{naoz2011}, and the
star's mass and rotational oblateness and the planet's mass  \citep{anderson_storch_lai2016,vick_lai_anderson2019}. 
It is worth revisiting these calculations to see if a better match to
the data can be obtained. A scenario we are investigating
is when the eccentricity required for high-eccentricity migration
can only be induced by an outer companion on a nearly polar orbit;
if the Jupiter begins its tidal migration near the general-relativistic quenching limit, the final obliquity would be near $90^\circ$. 

\item \textbf{Secular resonance crossing} was proposed by \cite{petrovich2020} to
explain the previously reported nearly-polar orbits of the Neptune-mass planets in our sample.
The resonance between the disk-driven nodal precession
frequencies of the transiting planet and an outer companion occurs
as the disk decreases in mass.
The resonance excites the inclination of the inner planet, and if the general relativistic
precession rate is fast enough, the inclination is pushed up to $90^\circ$.
Like the Von Zeipel-Kozai-Lidov scenario, the clearest prediction is that the nearly-polar systems
have massive outer companions.
\cite{petrovich2020} determined that the secular resonance crossing mechanism is most effective for lower-mass, close-orbiting
planets, and low-mass, slowly-rotating stars. Therefore, it is not clear that this
mechanism would work for many of the systems in our sample.

\item \textbf{Magnetic warping} can tilt the young proto-planetary disk toward a perpendicular orientation, but other mechanisms can counteract this effect (e.g., accretion, magnetic braking, disk winds, differential precession, and wrapping of magnetic fields around the stellar rotational axis; \citealt{foucart2011,lai2011,romanova2020}).
    
\end{enumerate}

Of course, nature is under no obligation to use only a single mechanism
to tilt orbits and stars. We may find that the systems in our sample
have followed different paths to their nearly perpendicular configurations.

\startlongtable
\begin{deluxetable*}{l LL LL LL L }
\tablecaption{Listing of the systems and some  parameters. \label{tab:systems}}
\tablewidth{0pt}
\tablehead{
\colhead{System} 
& T$_{\rm eff}$ & R & $P_\mathrm{rot}$ & $v \sin i$ & $\lambda$ & $\psi$& \mathrm{References} \\
\colhead{ }  & \mathrm{(K)} & (\mathrm{R}$_\odot$) & \mathrm{(days)} & \mathrm{(km\,s}$^{-1}$) & $(^{\circ})$ 
& $(^{\circ})$ &}
\decimalcolnumbers
\startdata
AU Mic & 3700\pm 100 & 0.75\pm0.03 & 4.85\pm0.75 & 9.23^{+0.79}_{-0.31} & 4.7^{+ 6.4}_{- 6.8} & 12.1^{+11.3}_{- 7.5} &  1 \\ 
CoRoT-2 & 5598\pm  50 & 0.90\pm0.02 & 4.52\pm0.02 & 11.25\pm0.45 & 1.0^{+ 7.7}_{- 6.0} & 8.9^{+ 6.7}_{- 5.1} &  2,3 \\ 
CoRoT-18 & 5440\pm 100 & 0.88^{+0.03}_{-0.03} & 5.53\pm0.33 & 8.00\pm1.00 & 10.0\pm20.0 & 25.4^{+13.6}_{-13.1} &  4 \\ 
DS Tuc & 5598^{+  28}_{-  59} & 0.87\pm0.03 & 2.85\pm0.06 & 20.58^{+0.31}_{-0.24} & 2.9^{+ 0.9}_{- 0.9} & 4.0^{+ 4.6}_{- 1.6} &  5 \\ 
EPIC 246851721 & 6202^{+  52}_{-  50} & 1.62^{+0.04}_{-0.04} & 1.14\pm0.06 & 74.92^{+0.62}_{-0.60} & 1.5\pm 0.9 & 2.9^{+10.8}_{- 1.6} &  6 \\ 
GJ 436 & 3416\pm  54 & 0.46\pm0.02 & 44.09\pm0.08 & 0.33^{+0.09}_{-0.07} & 72.0^{+33.0}_{-24.0} & 80.0^{+21.0}_{-18.0} &  7 \\ 
HAT-P-7 & 6310\pm  15 & 2.02\pm0.01 & -- & 2.70\pm0.50 & 142.0^{+12.0}_{-16.0} & 97.0\pm14.0 &  8,9,10 \\ 
HAT-P-11 & 4780\pm  50 & 0.68\pm0.01 & 29.32\pm1.00 & 1.00^{+0.95}_{-0.56} & 103.0^{+26.0}_{-10.0} & 97.0^{+ 8.0}_{- 4.0} &  11,12,13 \\ 
HAT-P-20 & 4595\pm  45 & 0.68\pm0.01 & 14.48\pm0.02 & 1.85\pm0.27 & 8.0\pm 6.9 & 36.9^{+12.4}_{-17.6} &  14 \\ 
HAT-P-22 & 5314\pm  50 & 1.06\pm0.05 & 28.70\pm0.40 & 1.65\pm0.26 & 2.1\pm 3.0 & 6.7^{+29.4}_{- 3.8} &  15 \\ 
HAT-P-36 & 5620\pm  40 & 1.04\pm0.02 & 15.30\pm0.40 & 3.12\pm0.75 & 14.0\pm18.0 & 28.6^{+16.0}_{-14.6} &  16 \\ 
HATS-2 & 5227\pm  95 & 0.90\pm0.02 & 24.98\pm0.04 & 1.50\pm0.50 & 8.0\pm 8.0 & 20.6^{+23.6}_{-10.4} &  2,17 \\ 
HD 63433 & 5640\pm  74 & 0.91^{+0.03}_{-0.03} & 6.61\pm0.71 & 7.30\pm0.30 & 8.0^{+33.0}_{-45.0} & 25.6^{+22.5}_{-15.3} &  18,19 \\ 
HD 189733 & 5050\pm  50 & 0.75\pm0.03 & 11.95\pm0.02 & 3.25\pm0.02 & 0.4\pm 0.2 & 2.3^{+13.5}_{- 1.6} &  20,21 \\ 
HD 209458 & 6117\pm  50 & 1.16\pm0.01 & 10.65\pm0.75 & 4.80\pm0.20 & 0.6\pm 0.4 & 28.2^{+ 9.7}_{-13.5} &  22,23 \\ 
K2-25 & 3207\pm  58 & 0.29\pm0.01 & 1.88\pm0.04 & 8.90\pm0.60 & 3.0\pm16.0 & 13.8^{+10.1}_{- 7.8} &  24 \\ 
K2-29 & 5358\pm  38 & 0.86\pm0.01 & 10.76\pm0.22 & 3.70\pm0.50 & 1.5\pm 8.7 & 19.3^{+13.7}_{-11.1} &  25 \\ 
K2-290 & 6302\pm 120 & 1.51^{+0.08}_{-0.08} & 6.63\pm0.66 & 6.90^{+0.50}_{-0.60} & 153.0\pm 8.0 & 124.0\pm 6.0 &  26,27 \\ 
KELT-9 & 9600\pm 400 & 2.42\pm0.06 & -- & 116.90\pm1.80 & 85.0\pm 0.2 & 87.0^{+10.0}_{-11.0} &  28,29 \\ 
KELT-17 & 7454\pm  49 & 1.65^{+0.06}_{-0.06} & -- & 44.20^{+1.50}_{-1.30} & 115.9\pm 4.1 & 116.0\pm 4.0 &  30 \\ 
Kepler-8 & 6213\pm 150 & 1.50\pm0.04 & 7.13\pm0.14 & 8.90\pm1.00 & 5.0\pm 7.0 & 31.1^{+10.9}_{-15.7} &  8 \\ 
Kepler-9 & 5774\pm  60 & 0.96\pm0.02 & 16.49\pm0.33 & 2.74\pm0.40 & 13.0\pm16.0 & 28.1^{+13.0}_{-13.6} &  31 \\ 
Kepler-13 & 7650\pm 250 & 1.71\pm0.04 & -- & 62.70\pm0.20 & 59.2\pm 0.1 & 60.2\pm 0.1 &  32 \\ 
Kepler-17 & 5781\pm  85 & 0.98^{+0.02}_{-0.05} & 12.09\pm0.24 & 4.70\pm1.00 & 0.0\pm15.0 & 19.7^{+14.4}_{-10.9} &  33 \\ 
Kepler-25 & 6270\pm  79 & 1.32^{+0.02}_{-0.01} & 23.15\pm0.04 & 8.20\pm0.20 & 0.5\pm 5.7 & 5.7^{+ 4.2}_{- 3.2} &  34,35 \\ 
Kepler-63 & 5576\pm  50 & 0.90^{+0.03}_{-0.02} & 5.40\pm0.01 & 5.60\pm0.80 & 110.0^{+14.0}_{-22.0} & 104.0^{+ 9.0}_{-14.0} &  22,36 \\ 
Kepler-448 & 6820\pm 120 & 1.63\pm0.15 & 1.29\pm0.03 & 66.43^{+1.00}_{-0.95} & 7.1^{+ 2.8}_{- 4.2} & 10.0^{+10.4}_{- 4.5} &  37 \\ 
MASCARA-4 & 7800\pm 200 & 1.92\pm0.11 & -- & 46.50\pm1.00 & 112.5^{+ 1.7}_{- 1.5} & 104.0^{+ 7.0}_{-13.0} &  38,39 \\ 
Qatar-1 & 4910\pm 100 & 0.80\pm0.02 & 23.70\pm0.12 & 1.70\pm0.30 & 8.4\pm 7.1 & 16.9^{+15.6}_{- 9.0} &  40,41 \\ 
Qatar-2 & 4645\pm  50 & 0.70\pm0.01 & 18.50\pm1.90 & 2.80\pm0.50 & 0.0\pm 8.0 & 11.4^{+12.3}_{- 7.1} &  42,43,44 \\ 
TRAPPIST-1 & 2557\pm  47 & 0.12\pm0.00 & 3.28\pm0.22 & 2.04\pm0.18 & 15.0^{+26.0}_{-30.0} & 23.3^{+17.0}_{-13.6} &  45 \\ 
WASP-4 & 5540\pm  55 & 0.91\pm0.02 & 22.20\pm3.30 & 2.14^{+0.38}_{-0.35} & 1.0^{+12.0}_{-14.0} & 20.0^{+15.2}_{-11.1} &  2,46 \\ 
WASP-5 & 5770\pm  65 & 1.09\pm0.04 & 16.20\pm0.40 & 3.20\pm0.30 & 12.1^{+ 8.0}_{-10.0} & 22.2^{+11.5}_{-10.3} &  2,47 \\ 
WASP-6 & 5375\pm  65 & 0.86\pm0.03 & 23.80\pm0.15 & 1.60^{+0.27}_{-0.17} & 7.2\pm 3.7 & 13.1^{+22.9}_{- 3.8} &  2,48,49 \\ 
WASP-7 & 6520\pm  70 & 1.48\pm0.09 & 3.68\pm1.23 & 14.00\pm2.00 & 86.0\pm 6.0 & 87.1^{+ 5.1}_{- 5.3} &  50 \\ 
WASP-8 & 5690\pm  36 & 0.98\pm0.02 & 15.31\pm0.80 & 1.90\pm0.05 & 143.0^{+ 1.5}_{- 1.6} & 118.2^{+ 3.2}_{- 3.0} &  51 \\ 
WASP-12 & 6313\pm  52 & 1.66^{+0.05}_{-0.04} & 6.77\pm1.58 & 1.60^{+0.80}_{-0.40} & 59.0^{+15.0}_{-20.0} & 85.5^{+ 6.8}_{- 7.8} &  8 \\ 
WASP-19 & 5460\pm  90 & 1.02\pm0.01 & 12.13\pm2.10 & 4.40\pm0.90 & 1.0\pm 1.2 & 3.7^{+28.0}_{- 2.3} &  8,52 \\ 
WASP-32 & 6100\pm 100 & 1.11\pm0.05 & 11.60\pm1.00 & 3.90^{+0.40}_{-0.50} & 10.5^{+ 6.4}_{- 6.5} & 35.8^{+10.2}_{-17.2} &  53,54 \\ 
WASP-33 & 7430\pm 100 & 1.51^{+0.02}_{-0.03} & 0.52\pm0.05 & 86.63^{+0.32}_{-0.37} & 112.9^{+ 0.2}_{- 0.2} & 104.1^{+ 2.8}_{- 2.8} &  55 \\ 
WASP-41 & 5546\pm  33 & 0.89\pm0.01 & 18.41\pm0.05 & 1.60\pm1.10 & 6.0\pm11.0 & 22.6^{+28.9}_{-11.9} &  2,56 \\ 
WASP-43 & 4520\pm 120 & 0.67^{+0.01}_{-0.01} & 15.60\pm0.40 & 2.26\pm0.54 & 3.5\pm 6.8 & 12.7^{+19.7}_{- 7.7} &  2,14 \\ 
WASP-52 & 5000\pm 100 & 0.79\pm0.02 & 17.26^{+0.51}_{-0.39} & 2.62\pm0.07 & 1.1\pm 1.1 & 2.8^{+ 9.8}_{- 1.8} &  57,58 \\ 
WASP-62 & 6230\pm  80 & 1.28\pm0.05 & 6.65\pm0.13 & 9.30\pm0.20 & 19.4^{+ 5.1}_{- 4.9} & 25.0^{+ 6.6}_{- 6.2} &  59 \\ 
WASP-69 & 4700\pm  50 & 0.81\pm0.03 & 23.07\pm0.16 & 2.20\pm0.40 & 0.4^{+ 2.0}_{- 1.9} & 3.8^{+18.5}_{- 2.7} &  2,60 \\ 
WASP-76 & 6329\pm  65 & 1.76\pm0.07 & 9.29\pm1.27 & 1.48\pm0.28 & 61.3^{+ 7.6}_{- 5.1} & 85.7^{+ 2.5}_{- 2.4} &  61 \\ 
WASP-84 & 5280\pm  80 & 0.77\pm0.02 & 14.36\pm0.35 & 2.56\pm0.08 & 0.3\pm 1.7 & 18.6^{+ 4.4}_{-15.7} &  2,62 \\ 
WASP-85 & 5685\pm  65 & 0.94\pm0.02 & 13.08\pm0.26 & 3.41\pm0.89 & 0.0\pm14.0 & 22.1^{+17.4}_{-12.1} &  63 \\ 
WASP-94A & 6170\pm  80 & 1.62^{+0.05}_{-0.04} & 10.48\pm1.60 & 4.20\pm0.50 & 151.0^{+16.0}_{-23.0} & 116.6^{+ 9.9}_{- 9.1} &  64 \\ 
WASP-107 & 4425\pm  70 & 0.67\pm0.02 & 17.10\pm1.00 & 2.50\pm0.80 & 112.6^{+24.9}_{-20.6} & 92.6^{+30.7}_{- 1.8} &  65,66 \\ 
WASP-121 & 6586\pm  59 & 1.44\pm0.03 & 3.38\pm0.40 & 13.56^{+0.69}_{-0.68} & 87.2^{+ 0.4}_{- 0.4} & 88.1\pm 0.2 &  67 \\ 
WASP-166 & 6050\pm  50 & 1.22\pm0.06 & 12.30\pm1.90 & 5.10\pm0.30 & 3.0\pm 5.0 & 10.1^{+16.7}_{- 5.9} &  68 \\ 
WASP-167 & 7043^{+  89}_{-  68} & 1.79\pm0.05 & 1.02\pm0.10 & 49.94\pm0.04 & 165.0\pm 5.0 & 123.8^{+11.6}_{-10.6} &  69 \\ 
WASP-189 & 8000\pm  80 & 2.36\pm0.03 & -- & 100.00\pm5.00 & 89.3\pm 1.4 & 85.4\pm 4.3 &  70,71 \\ 
XO-2 & 5332\pm  57 & 1.00^{+0.03}_{-0.03} & 41.60\pm1.10 & 1.07\pm0.09 & 7.0\pm11.0 & 26.5^{+11.8}_{-13.7} &  72 \\ 
XO-6 & 6720\pm 100 & 1.93\pm0.18 & 1.79\pm0.06 & 48.00\pm3.00 & 20.7\pm 2.3 & 23.3^{+14.3}_{- 2.9} &  73 \\ 
pi Men & 5998\pm  62 & 1.17\pm0.02 & 18.30\pm1.00 & 3.16\pm0.27 & 24.0\pm 4.1 & 26.9^{+ 5.8}_{- 4.7} &  74,75 \\ 
\enddata
\tablecomments{The quoted $1-\sigma$ values for for P$_\mathrm{rot}$ and $\psi$ are obtained from the Highest Density Regions. If not derived in this work then values are either taken from \href{https://www.astro.keele.ac.uk/jkt/tepcat/}{TEPCAT} or:
 1 \citep{Hirano+2020}, 2 \citep{Bonomo+2017}, 3 \citep{Czesla+2012}, 4 \citep{Hebrard+2011}, 5 \citep{Zhou+2020}, 6 \citep{Yu+2018}, 7 \citep{Bourrier+2018}, 8 \citep{AlbrechtWinnJohnson+2012}, 9 \citep{Masuda2015}, 10 \citep{Lund+2014}, 11 \citep{Beky+2014}, 12 \citep{Winn+2010_hatp11}, 13 \citep{Sanchis-OjedaWinn2011}, 14 \citep{Esposito+2017}, 15 \citep{Mancini+2018}, 16 \citep{Mancini+2015b}, 17 \citep{Mohler-Fischer+2013}, 18 \citep{Mann+2020}, 19 \citep{Dai+2020}, 20 \citep{HenryWinn2008}, 21 \citep{Cegla+2016}, 22 \citep{MaxtedSerenelliSouthworth2015}, 23 \citep{Santos+2020}, 24 \citep{Stefansson+2020}, 25 \citep{Santerne+2016}, 26 \citep{Hjorth+2019}, 27 \citep{Hjorth+2021}, 28 \citep{Wyttenbach+2020}, 29 \citep{Ahlers+2020b}, 30 \citep{Zhou+2016b}, 31 \citep{Wang+2018}, 32 \citep{HowarthMorello2017}, 33 \citep{Desert+2011}, 34 \citep{McQuillanMazehAigrain2013}, 35 \citep{AlbrechtWinnMacrcy+2013}, 36 \citep{Sanchis-Ojeda+2013}, 37 \citep{Johnson+2017}, 38 \citep{Dorval+2020}, 39 \citep{Ahlers+2020}, 40 \citep{Mislis+2015}, 41 \citep{Covino+2013}, 42 \citep{Dai+2017}, 43 \citep{Bryan+2012}, 44 \citep{MocnikSouthworthHellier2017}, 45 \citep{Hirano+2020b}, 46 \citep{Sanchis-Ojeda+2011}, 47 \citep{Triaud+2010}, 48 \citep{Gillion+2009}, 49 \citep{Tregloan-Reed+2015}, 50 \citep{AlbrechtWinnButler+2012}, 51 \citep{Bourrier+2017}, 52 \citep{Tregloan-reedSouthworthTappert2012}, 53 \citep{Brothwell+2014}, 54 \citep{Brown+2012b}, 55 \citep{Johnson+2015}, 56 \citep{Southworth+2016}, 57 \citep{Rosich+2020}, 58 \citep{Chen+2020}, 59 \citep{Brown+2017}, 60 \citep{Casasayas-Barris+2017}, 61 \citep{Ehrenreich2020}, 62 \citep{Anderson+2015}, 63 \citep{Mocnik+2016}, 64 \citep{Neveu-VanMalle+2014}, 65 \citep{Anderson+2017}, 66 \citep{Rubenzahl+2021}, 67 \citep{Bourrier+2020}, 68 \citep{Hellier+2019}, 69 \citep{Temple+2017}, 70 \citep{Anderson+2018}, 71 \citep{Lendl+2020}, 72 \citep{Damasso+2015}, 73 \citep{Crouzet+2017}, 74 \citep{Zurlo+2018}, 75 \citep{Kunovac_-Hodzic+2021}
}
\end{deluxetable*}

\begin{acknowledgments}
We thank R.\ Rubenzahl for providing posterior
samples for $\lambda$ and $\psi$ based on his
group's analysis of WASP-107. JNW thanks Kassandra Anderson and
Scott Tremaine for interesting discussions.
Funding for the Stellar Astrophysics Centre is provided by The Danish National Research Foundation (Grant agreement no.: DNRF106). Work by JNW was partly supported by the Heising-Simons Foundation. RID is supported in part by NASA XRP NNX16AB50G. This paper includes data collected by the \textit{TESS} mission, which was funded by the NASA Science Mission directorate. This research has made use of the  \href{https://exoplanetarchive.ipac.caltech.edu}{NASA Exoplanet Archive}, which is operated by the California Institute of Technology, under contract with the National Aeronautics and Space Administration under the Exoplanet Exploration Program. \end{acknowledgments}
 
\vspace{5mm}

\facilities{TESS}

\software{Matplotlib \citep{Hunter:2007}, astropy \citep{astropy:2013,astropy:2018}, spotify}


\begin{thebibliography}{}
\expandafter\ifx\csname natexlab\endcsname\relax\def\natexlab#1{#1}\fi
\providecommand{\url}[1]{\href{#1}{#1}}
\providecommand{\dodoi}[1]{doi:~\href{http://doi.org/#1}{\nolinkurl{#1}}}
\providecommand{\doeprint}[1]{\href{http://ascl.net/#1}{\nolinkurl{http://ascl.net/#1}}}
\providecommand{\doarXiv}[1]{\href{https://arxiv.org/abs/#1}{\nolinkurl{https://arxiv.org/abs/#1}}}

\bibitem[{{Ahlers} {et~al.}(2020{\natexlab{a}}){Ahlers}, {Johnson}, {Stassun},
  {Col{\'o}n}, {Barnes}, {Stevens}, {Beatty}, {Gaudi}, {Collins}, {Rodriguez},
  {Ricker}, {Vanderspek}, {Latham}, {Seager}, {Winn}, {Jenkins}, {Caldwell},
  {Goeke}, {Osborn}, {Paegert}, {Rowden}, \& {Tenenbaum}}]{Ahlers+2020b}
{Ahlers}, J.~P., {Johnson}, M.~C., {Stassun}, K.~G., {et~al.}
  2020{\natexlab{a}}, \aj, 160, 4, \dodoi{10.3847/1538-3881/ab8fa3}

\bibitem[{{Ahlers} {et~al.}(2020{\natexlab{b}}){Ahlers}, {Kruse}, {Col{\'o}n},
  {Dorval}, {Talens}, {Snellen}, {Albrecht}, {Otten}, {Ricker}, {Vand erspek},
  {Latham}, {Seager}, {Winn}, {Jenkins}, {Haworth}, {Cartwright}, {Morris},
  {Rowden}, {Tenenbaum}, \& {Ting}}]{Ahlers+2020}
{Ahlers}, J.~P., {Kruse}, E., {Col{\'o}n}, K.~D., {et~al.} 2020{\natexlab{b}},
  \apj, 888, 63, \dodoi{10.3847/1538-4357/ab59d0}

\bibitem[{{Albrecht} {et~al.}(2012{\natexlab{a}}){Albrecht}, {Winn}, {Butler},
  {Crane}, {Shectman}, {Thompson}, {Hirano}, \&
  {Wittenmyer}}]{AlbrechtWinnButler+2012}
{Albrecht}, S., {Winn}, J.~N., {Butler}, R.~P., {et~al.} 2012{\natexlab{a}},
  \apj, 744, 189, \dodoi{10.1088/0004-637X/744/2/189}

\bibitem[{{Albrecht} {et~al.}(2013){Albrecht}, {Winn}, {Marcy}, {Howard},
  {Isaacson}, \& {Johnson}}]{AlbrechtWinnMacrcy+2013}
{Albrecht}, S., {Winn}, J.~N., {Marcy}, G.~W., {et~al.} 2013, \apj, 771, 11,
  \dodoi{10.1088/0004-637X/771/1/11}

\bibitem[{{Albrecht} {et~al.}(2012{\natexlab{b}}){Albrecht}, {Winn}, {Johnson},
  {Howard}, {Marcy}, {Butler}, {Arriagada}, {Crane}, {Shectman}, {Thompson},
  {Hirano}, {Bakos}, \& {Hartman}}]{AlbrechtWinnJohnson+2012}
{Albrecht}, S., {Winn}, J.~N., {Johnson}, J.~A., {et~al.} 2012{\natexlab{b}},
  \apj, 757, \dodoi{10.1088/0004-637X/757/1/18}

\bibitem[{{Anderson} {et~al.}(2015){Anderson}, {Triaud}, {Turner}, {Brown},
  {Clark}, {Smalley}, {Collier Cameron}, {Doyle}, {Gillon}, {Hellier}, {Lovis},
  {Maxted}, {Pollacco}, {Queloz}, \& {Smith}}]{Anderson+2015}
{Anderson}, D.~R., {Triaud}, A.~H.~M.~J., {Turner}, O.~D., {et~al.} 2015,
  \apjl, 800, L9, \dodoi{10.1088/2041-8205/800/1/L9}

\bibitem[{{Anderson} {et~al.}(2017){Anderson}, {Collier Cameron}, {Delrez},
  {Doyle}, {Gillon}, {Hellier}, {Jehin}, {Lendl}, {Maxted}, {Madhusudhan},
  {Pepe}, {Pollacco}, {Queloz}, {S{\'e}gransan}, {Smalley}, {Smith}, {Triaud},
  {Turner}, {Udry}, \& {West}}]{Anderson+2017}
{Anderson}, D.~R., {Collier Cameron}, A., {Delrez}, L., {et~al.} 2017, \aap,
  604, A110, \dodoi{10.1051/0004-6361/201730439}

\bibitem[{{Anderson} {et~al.}(2018){Anderson}, {Temple}, {Nielsen}, {Burdanov},
  {Hellier}, {Bouchy}, {Brown}, {Collier Cameron}, {Gillon}, {Jehin}, {Maxted},
  {Pepe}, {Pollacco}, {Pozuelos}, {Queloz}, {S{\'e}gransan}, {Smalley},
  {Triaud}, {Turner}, {Udry}, \& {West}}]{Anderson+2018}
{Anderson}, D.~R., {Temple}, L.~Y., {Nielsen}, L.~D., {et~al.} 2018, arXiv
  e-prints, arXiv:1809.04897.
\newblock \doarXiv{1809.04897}

\bibitem[{{Anderson} {et~al.}(2016){Anderson}, {Storch}, \&
  {Lai}}]{anderson_storch_lai2016}
{Anderson}, K.~R., {Storch}, N.~I., \& {Lai}, D. 2016, \mnras, 456, 3671,
  \dodoi{10.1093/mnras/stv2906}

\bibitem[{{Anderson} {et~al.}(2021){Anderson}, {Winn}, \&
  {Penev}}]{anderson2021}
{Anderson}, K.~R., {Winn}, J.~N., \& {Penev}, K. 2021, arXiv e-prints,
  arXiv:2102.01081.
\newblock \doarXiv{2102.01081}

\bibitem[{{Astropy Collaboration} {et~al.}(2013){Astropy Collaboration},
  {Robitaille}, {Tollerud}, {Greenfield}, {Droettboom}, {Bray}, {Aldcroft},
  {Davis}, {Ginsburg}, {Price-Whelan}, {Kerzendorf}, {Conley}, {Crighton},
  {Barbary}, {Muna}, {Ferguson}, {Grollier}, {Parikh}, {Nair}, {Unther},
  {Deil}, {Woillez}, {Conseil}, {Kramer}, {Turner}, {Singer}, {Fox}, {Weaver},
  {Zabalza}, {Edwards}, {Azalee Bostroem}, {Burke}, {Casey}, {Crawford},
  {Dencheva}, {Ely}, {Jenness}, {Labrie}, {Lim}, {Pierfederici}, {Pontzen},
  {Ptak}, {Refsdal}, {Servillat}, \& {Streicher}}]{astropy:2013}
{Astropy Collaboration}, {Robitaille}, T.~P., {Tollerud}, E.~J., {et~al.} 2013,
  \aap, 558, A33, \dodoi{10.1051/0004-6361/201322068}

\bibitem[{{Astropy Collaboration} {et~al.}(2018){Astropy Collaboration},
  {Price-Whelan}, {Sip{\H{o}}cz}, {G{\"u}nther}, {Lim}, {Crawford}, {Conseil},
  {Shupe}, {Craig}, {Dencheva}, {Ginsburg}, {Vand erPlas}, {Bradley},
  {P{\'e}rez-Su{\'a}rez}, {de Val-Borro}, {Aldcroft}, {Cruz}, {Robitaille},
  {Tollerud}, {Ardelean}, {Babej}, {Bach}, {Bachetti}, {Bakanov}, {Bamford},
  {Barentsen}, {Barmby}, {Baumbach}, {Berry}, {Biscani}, {Boquien}, {Bostroem},
  {Bouma}, {Brammer}, {Bray}, {Breytenbach}, {Buddelmeijer}, {Burke},
  {Calderone}, {Cano Rodr{\'\i}guez}, {Cara}, {Cardoso}, {Cheedella}, {Copin},
  {Corrales}, {Crichton}, {D'Avella}, {Deil}, {Depagne}, {Dietrich}, {Donath},
  {Droettboom}, {Earl}, {Erben}, {Fabbro}, {Ferreira}, {Finethy}, {Fox},
  {Garrison}, {Gibbons}, {Goldstein}, {Gommers}, {Greco}, {Greenfield},
  {Groener}, {Grollier}, {Hagen}, {Hirst}, {Homeier}, {Horton}, {Hosseinzadeh},
  {Hu}, {Hunkeler}, {Ivezi{\'c}}, {Jain}, {Jenness}, {Kanarek}, {Kendrew},
  {Kern}, {Kerzendorf}, {Khvalko}, {King}, {Kirkby}, {Kulkarni}, {Kumar},
  {Lee}, {Lenz}, {Littlefair}, {Ma}, {Macleod}, {Mastropietro}, {McCully},
  {Montagnac}, {Morris}, {Mueller}, {Mumford}, {Muna}, {Murphy}, {Nelson},
  {Nguyen}, {Ninan}, {N{\"o}the}, {Ogaz}, {Oh}, {Parejko}, {Parley}, {Pascual},
  {Patil}, {Patil}, {Plunkett}, {Prochaska}, {Rastogi}, {Reddy Janga},
  {Sabater}, {Sakurikar}, {Seifert}, {Sherbert}, {Sherwood-Taylor}, {Shih},
  {Sick}, {Silbiger}, {Singanamalla}, {Singer}, {Sladen}, {Sooley},
  {Sornarajah}, {Streicher}, {Teuben}, {Thomas}, {Tremblay}, {Turner},
  {Terr{\'o}n}, {van Kerkwijk}, {de la Vega}, {Watkins}, {Weaver}, {Whitmore},
  {Woillez}, {Zabalza}, \& {Astropy Contributors}}]{astropy:2018}
{Astropy Collaboration}, {Price-Whelan}, A.~M., {Sip{\H{o}}cz}, B.~M., {et~al.}
  2018, \aj, 156, 123, \dodoi{10.3847/1538-3881/aabc4f}

\bibitem[{{B{\'e}ky} {et~al.}(2014){B{\'e}ky}, {Holman}, {Kipping}, \&
  {Noyes}}]{Beky+2014}
{B{\'e}ky}, B., {Holman}, M.~J., {Kipping}, D.~M., \& {Noyes}, R.~W. 2014,
  \apj, 788, 1, \dodoi{10.1088/0004-637X/788/1/1}

\bibitem[{{Bonomo} {et~al.}(2017){Bonomo}, {Desidera}, {Benatti}, {Borsa},
  {Crespi}, {Damasso}, {Lanza}, {Sozzetti}, {Lodato}, {Marzari}, {Boccato},
  {Claudi}, {Cosentino}, {Covino}, {Gratton}, {Maggio}, {Micela}, {Molinari},
  {Pagano}, {Piotto}, {Poretti}, {Smareglia}, {Affer}, {Biazzo}, {Bignamini},
  {Esposito}, {Giacobbe}, {H{\'e}brard}, {Malavolta}, {Maldonado}, {Mancini},
  {Martinez Fiorenzano}, {Masiero}, {Nascimbeni}, {Pedani}, {Rainer}, \&
  {Scandariato}}]{Bonomo+2017}
{Bonomo}, A.~S., {Desidera}, S., {Benatti}, S., {et~al.} 2017, \aap, 602, A107,
  \dodoi{10.1051/0004-6361/201629882}

\bibitem[{{Bourrier} {et~al.}(2017){Bourrier}, {Cegla}, {Lovis}, \&
  {Wyttenbach}}]{Bourrier+2017}
{Bourrier}, V., {Cegla}, H.~M., {Lovis}, C., \& {Wyttenbach}, A. 2017, \aap,
  599, A33, \dodoi{10.1051/0004-6361/201629973}

\bibitem[{{Bourrier} {et~al.}(2018){Bourrier}, {Lovis}, {Beust}, {Ehrenreich},
  {Henry}, {Astudillo-Defru}, {Allart}, {Bonfils}, {S{\'e}gransan}, {Delfosse},
  {Cegla}, {Wyttenbach}, {Heng}, {Lavie}, \& {Pepe}}]{Bourrier+2018}
{Bourrier}, V., {Lovis}, C., {Beust}, H., {et~al.} 2018, \nat, 553, 477,
  \dodoi{10.1038/nature24677}

\bibitem[{{Bourrier} {et~al.}(2020){Bourrier}, {Ehrenreich}, {Lendl},
  {Cretignier}, {Allart}, {Dumusque}, {Cegla}, {Su{\'a}rez-Mascare{\~n}o},
  {Wyttenbach}, {Hoeijmakers}, {Melo}, {Kuntzer}, {Astudillo-Defru}, {Giles},
  {Heng}, {Kitzmann}, {Lavie}, {Lovis}, {Murgas}, {Nascimbeni}, {Pepe}, {Pino},
  {Segransan}, \& {Udry}}]{Bourrier+2020}
{Bourrier}, V., {Ehrenreich}, D., {Lendl}, M., {et~al.} 2020, \aap, 635, A205,
  \dodoi{10.1051/0004-6361/201936640}

\bibitem[{{Brothwell} {et~al.}(2014){Brothwell}, {Watson}, {H{\'e}brard},
  {Triaud}, {Cegla}, {Santerne}, {H{\'e}brard}, {Anderson}, {Pollacco},
  {Simpson}, {Bouchy}, {Brown}, {Chew}, {Collier Cameron}, {Armstrong},
  {Barros}, {Bento}, {Bochinski}, {Burwitz}, {Busuttil}, {Delrez}, {Doyle},
  {Faedi}, {Fumel}, {Gillon}, {Haswell}, {Hellier}, {Jehin}, {Kolb}, {Lendl},
  {Liebig}, {Maxted}, {McCormac}, {Miller}, {Norton}, {Pepe}, {Queloz},
  {Rodr{\'\i}guez}, {S{\'e}gransan}, {Skillen}, {Smalley}, {Stassun}, {Udry},
  {West}, \& {Wheatley}}]{Brothwell+2014}
{Brothwell}, R.~D., {Watson}, C.~A., {H{\'e}brard}, G., {et~al.} 2014, \mnras,
  440, 3392, \dodoi{10.1093/mnras/stu520}

\bibitem[{{Brown} {et~al.}(2012){Brown}, {Collier Cameron}, {D{\'{\i}}az},
  {Doyle}, {Gillon}, {Lendl}, {Smalley}, {Triaud}, {Anderson}, {Enoch},
  {Hellier}, {Maxted}, {Miller}, {Pollacco}, {Queloz}, {Boisse}, \&
  {H{\'e}brard}}]{Brown+2012b}
{Brown}, D.~J.~A., {Collier Cameron}, A., {D{\'{\i}}az}, R.~F., {et~al.} 2012,
  \apj, 760, 139, \dodoi{10.1088/0004-637X/760/2/139}

\bibitem[{{Brown} {et~al.}(2017){Brown}, {Triaud}, {Doyle}, {Gillon}, {Lendl},
  {Anderson}, {Collier Cameron}, {H{\'e}brard}, {Hellier}, {Lovis}, {Maxted},
  {Pepe}, {Pollacco}, {Queloz}, \& {Smalley}}]{Brown+2017}
{Brown}, D.~J.~A., {Triaud}, A.~H.~M.~J., {Doyle}, A.~P., {et~al.} 2017,
  \mnras, 464, 810, \dodoi{10.1093/mnras/stw2316}

\bibitem[{{Bryan} {et~al.}(2012){Bryan}, {Alsubai}, {Latham}, {Parley},
  {Collier Cameron}, {Quinn}, {Carter}, {Fulton}, {Berlind}, {Brown},
  {Buchhave}, {Calkins}, {Esquerdo}, {F{\H{u}}r{\'e}sz}, {Gr{\r{a}}e
  J{\o}rgensen}, {Horne}, {Stefanik}, {Street}, {Torres}, {West}, {Dominik},
  {Harps{\o}e}, {Liebig}, {Calchi Novati}, {Ricci}, \&
  {Skottfelt}}]{Bryan+2012}
{Bryan}, M.~L., {Alsubai}, K.~A., {Latham}, D.~W., {et~al.} 2012, \apj, 750,
  84, \dodoi{10.1088/0004-637X/750/1/84}

\bibitem[{{Casasayas-Barris} {et~al.}(2017){Casasayas-Barris}, {Palle},
  {Nowak}, {Yan}, {Nortmann}, \& {Murgas}}]{Casasayas-Barris+2017}
{Casasayas-Barris}, N., {Palle}, E., {Nowak}, G., {et~al.} 2017, \aap, 608,
  A135, \dodoi{10.1051/0004-6361/201731956}

\bibitem[{{Cegla} {et~al.}(2016){Cegla}, {Lovis}, {Bourrier}, {Beeck},
  {Watson}, \& {Pepe}}]{Cegla+2016}
{Cegla}, H.~M., {Lovis}, C., {Bourrier}, V., {et~al.} 2016, \aap, 588, A127,
  \dodoi{10.1051/0004-6361/201527794}

\bibitem[{{Chen} {et~al.}(2020){Chen}, {Casasayas-Barris}, {Pall{\'e}}, {Yan},
  {Stangret}, {Cegla}, {Allart}, \& {Lovis}}]{Chen+2020}
{Chen}, G., {Casasayas-Barris}, N., {Pall{\'e}}, E., {et~al.} 2020, \aap, 635,
  A171, \dodoi{10.1051/0004-6361/201936986}

\bibitem[{{Covino} {et~al.}(2013){Covino}, {Esposito}, {Barbieri}, {Mancini},
  {Nascimbeni}, {Claudi}, {Desidera}, {Gratton}, {Lanza}, {Sozzetti}, {Biazzo},
  {Affer}, {Gandolfi}, {Munari}, {Pagano}, {Bonomo}, {Collier Cameron},
  {H{\'e}brard}, {Maggio}, {Messina}, {Micela}, {Molinari}, {Pepe}, {Piotto},
  {Ribas}, {Santos}, {Southworth}, {Shkolnik}, {Triaud}, {Bedin}, {Benatti},
  {Boccato}, {Bonavita}, {Borsa}, {Borsato}, {Brown}, {Carolo}, {Ciceri},
  {Cosentino}, {Damasso}, {Faedi}, {Mart{\'\i}nez Fiorenzano}, {Latham},
  {Lovis}, {Mordasini}, {Nikolov}, {Poretti}, {Rainer}, {Rebolo L{\'o}pez},
  {Scandariato}, {Silvotti}, {Smareglia}, {Alcal{\'a}}, {Cunial}, {Di
  Fabrizio}, {Di Mauro}, {Giacobbe}, {Granata}, {Harutyunyan}, {Knapic},
  {Lattanzi}, {Leto}, {Lodato}, {Malavolta}, {Marzari}, {Molinaro},
  {Nardiello}, {Pedani}, {Prisinzano}, \& {Turrini}}]{Covino+2013}
{Covino}, E., {Esposito}, M., {Barbieri}, M., {et~al.} 2013, \aap, 554, A28,
  \dodoi{10.1051/0004-6361/201321298}

\bibitem[{{Crouzet} {et~al.}(2017){Crouzet}, {McCullough}, {Long}, {Montanes
  Rodriguez}, {Lecavelier des Etangs}, {Ribas}, {Bourrier}, {H{\'e}brard},
  {Vilardell}, {Deleuil}, {Herrero}, {Garcia-Melendo}, {Akhenak}, {Foote},
  {Gary}, {Benni}, {Guillot}, {Conjat}, {M{\'e}karnia}, {Garlitz}, {Burke},
  {Courcol}, \& {Demangeon}}]{Crouzet+2017}
{Crouzet}, N., {McCullough}, P.~R., {Long}, D., {et~al.} 2017, \aj, 153, 94,
  \dodoi{10.3847/1538-3881/153/3/94}

\bibitem[{{Czesla} {et~al.}(2012){Czesla}, {Schr{\"o}ter}, {Wolter}, {von
  Essen}, {Huber}, {Schmitt}, {Reichart}, \& {Moore}}]{Czesla+2012}
{Czesla}, S., {Schr{\"o}ter}, S., {Wolter}, U., {et~al.} 2012, \aap, 539, A150,
  \dodoi{10.1051/0004-6361/201118042}

\bibitem[{{Dai} {et~al.}(2017){Dai}, {Winn}, {Yu}, \& {Albrecht}}]{Dai+2017}
{Dai}, F., {Winn}, J.~N., {Yu}, L., \& {Albrecht}, S. 2017, \aj, 153, 40,
  \dodoi{10.3847/1538-3881/153/1/40}

\bibitem[{{Dai} {et~al.}(2020){Dai}, {Roy}, {Fulton}, {Robertson}, {Hirsch},
  {Isaacson}, {Albrecht}, {Mann}, {Kristiansen}, {Batalha}, {Beard}, {Behmard},
  {Chontos}, {Crossfield}, {Dalba}, {Dressing}, {Giacalone}, {Hill}, {Howard},
  {Huber}, {Kane}, {Kosiarek}, {Lubin}, {Mayo}, {Mocnik}, {Akana Murphy},
  {Petigura}, {Rosenthal}, {Rubenzahl}, {Scarsdale}, {Weiss}, {Van Zandt},
  {Ricker}, {Vanderspek}, {Latham}, {Seager}, {Winn}, {Jenkins}, {Caldwell},
  {Charbonneau}, {Daylan}, {G{\"u}nther}, {Morgan}, {Quinn}, {Rose}, \&
  {Smith}}]{Dai+2020}
{Dai}, F., {Roy}, A., {Fulton}, B., {et~al.} 2020, arXiv e-prints,
  arXiv:2008.12397.
\newblock \doarXiv{2008.12397}

\bibitem[{{Damasso} {et~al.}(2015){Damasso}, {Biazzo}, {Bonomo}, {Desidera},
  {Lanza}, {Nascimbeni}, {Esposito}, {Scandariato}, {Sozzetti}, {Cosentino},
  {Gratton}, {Malavolta}, {Rainer}, {Gandolfi}, {Poretti}, {Zanmar Sanchez},
  {Ribas}, {Santos}, {Affer}, {Andreuzzi}, {Barbieri}, {Bedin}, {Benatti},
  {Bernagozzi}, {Bertolini}, {Bonavita}, {Borsa}, {Borsato}, {Boschin},
  {Calcidese}, {Carbognani}, {Cenadelli}, {Christille}, {Claudi}, {Covino},
  {Cunial}, {Giacobbe}, {Granata}, {Harutyunyan}, {Lattanzi}, {Leto},
  {Libralato}, {Lodato}, {Lorenzi}, {Mancini}, {Martinez Fiorenzano},
  {Marzari}, {Masiero}, {Micela}, {Molinari}, {Molinaro}, {Munari}, {Murabito},
  {Pagano}, {Pedani}, {Piotto}, {Rosenberg}, {Silvotti}, \&
  {Southworth}}]{Damasso+2015}
{Damasso}, M., {Biazzo}, K., {Bonomo}, A.~S., {et~al.} 2015, \aap, 575, A111,
  \dodoi{10.1051/0004-6361/201425332}

\bibitem[{{D{\'e}sert} {et~al.}(2011){D{\'e}sert}, {Charbonneau}, {Demory},
  {Ballard}, {Carter}, {Fortney}, {Cochran}, {Endl}, {Quinn}, {Isaacson},
  {Fressin}, {Buchhave}, {Latham}, {Knutson}, {Bryson}, {Torres}, {Rowe},
  {Batalha}, {Borucki}, {Brown}, {Caldwell}, {Christiansen}, {Deming},
  {Fabrycky}, {Ford}, {Gilliland}, {Gillon}, {Haas}, {Jenkins}, {Kinemuchi},
  {Koch}, {Lissauer}, {Lucas}, {Mullally}, {MacQueen}, {Marcy}, {Sasselov},
  {Seager}, {Still}, {Tenenbaum}, {Uddin}, \& {Winn}}]{Desert+2011}
{D{\'e}sert}, J.-M., {Charbonneau}, D., {Demory}, B.-O., {et~al.} 2011, \apjs,
  197, 14, \dodoi{10.1088/0067-0049/197/1/14}

\bibitem[{{Dorval} {et~al.}(2020){Dorval}, {Talens}, {Otten}, {Brahm},
  {Jord{\'a}n}, {Torres}, {Vanzi}, {Zapata}, {Henry}, {Paredes}, {Jao},
  {James}, {Hinojosa}, {Bakos}, {Csubry}, {Bhatti}, {Suc}, {Osip}, {Mamajek},
  {Mellon}, {Wyttenbach}, {Stuik}, {Kenworthy}, {Bailey}, {Ireland},
  {Crawford}, {Lomberg}, {Kuhn}, \& {Snellen}}]{Dorval+2020}
{Dorval}, P., {Talens}, G.~J.~J., {Otten}, G.~P.~P.~L., {et~al.} 2020, \aap,
  635, A60, \dodoi{10.1051/0004-6361/201935611}

\bibitem[{{Ehrenreich} {et~al.}(2020){Ehrenreich}, {Lovis}, {Allart}, {Zapatero
  Osorio}, {Pepe}, {Cristiani}, {Rebolo}, {Santos}, {Borsa}, {Demangeon},
  {Dumusque}, {Gonz{\'a}lez Hern{\'a}ndez}, {Casasayas-Barris},
  {S{\'e}gransan}, {Sousa}, {Abreu}, {Adibekyan}, {Affolter}, {Allende Prieto},
  {Alibert}, {Aliverti}, {Alves}, {Amate}, {Avila}, {Baldini}, {Bandy}, {Benz},
  {Bianco}, {Bolmont}, {Bouchy}, {Bourrier}, {Broeg}, {Cabral}, {Calderone},
  {Pall{\'e}}, {Cegla}, {Cirami}, {Coelho}, {Conconi}, {Coretti}, {Cumani},
  {Cupani}, {Dekker}, {Delabre}, {Deiries}, {D'Odorico}, {Di Marcantonio},
  {Figueira}, {Fragoso}, {Genolet}, {Genoni}, {G{\'e}nova Santos}, {Hara},
  {Hughes}, {Iwert}, {Kerber}, {Knudstrup}, {Landoni}, {Lavie}, {Lizon},
  {Lendl}, {Lo Curto}, {Maire}, {Manescau}, {Martins}, {M{\'e}gevand},
  {Mehner}, {Micela}, {Modigliani}, {Molaro}, {Monteiro}, {Monteiro},
  {Moschetti}, {M{\"u}ller}, {Nunes}, {Oggioni}, {Oliveira}, {Pariani},
  {Pasquini}, {Poretti}, {Rasilla}, {Redaelli}, {Riva}, {Santana Tschudi},
  {Santin}, {Santos}, {Segovia Milla}, {Seidel}, {Sosnowska}, {Sozzetti},
  {Span{\`o}}, {Su{\'a}rez Mascare{\~n}o}, {Tabernero}, {Tenegi}, {Udry},
  {Zanutta}, \& {Zerbi}}]{Ehrenreich2020}
{Ehrenreich}, D., {Lovis}, C., {Allart}, R., {et~al.} 2020, \nat, 580, 597,
  \dodoi{10.1038/s41586-020-2107-1}

\bibitem[{{Esposito} {et~al.}(2017){Esposito}, {Covino}, {Desidera}, {Mancini},
  {Nascimbeni}, {Zanmar Sanchez}, {Biazzo}, {Lanza}, {Leto}, {Southworth},
  {Bonomo}, {Su{\'a}rez Mascare{\~n}o}, {Boccato}, {Cosentino}, {Claudi},
  {Gratton}, {Maggio}, {Micela}, {Molinari}, {Pagano}, {Piotto}, {Poretti},
  {Smareglia}, {Sozzetti}, {Affer}, {Anderson}, {Andreuzzi}, {Benatti},
  {Bignamini}, {Borsa}, {Borsato}, {Ciceri}, {Damasso}, {di Fabrizio},
  {Giacobbe}, {Granata}, {Harutyunyan}, {Henning}, {Malavolta}, {Maldonado},
  {Martinez Fiorenzano}, {Masiero}, {Molaro}, {Molinaro}, {Pedani}, {Rainer},
  {Scandariato}, \& {Turner}}]{Esposito+2017}
{Esposito}, M., {Covino}, E., {Desidera}, S., {et~al.} 2017, \aap, 601, A53,
  \dodoi{10.1051/0004-6361/201629720}

\bibitem[{{Fabrycky} \& {Tremaine}(2007)}]{fabrycky2007}
{Fabrycky}, D., \& {Tremaine}, S. 2007, \apj, 669, 1298, \dodoi{10.1086/521702}

\bibitem[{{Fabrycky} \& {Winn}(2009)}]{fabrycky_winn2009}
{Fabrycky}, D.~C., \& {Winn}, J.~N. 2009, \apj, 696, 1230,
  \dodoi{10.1088/0004-637X/696/2/1230}

\bibitem[{{Foucart} \& {Lai}(2011)}]{foucart2011}
{Foucart}, F., \& {Lai}, D. 2011, \mnras, 412, 2799,
  \dodoi{10.1111/j.1365-2966.2010.18176.x}

\bibitem[{{Gillon} {et~al.}(2009){Gillon}, {Anderson}, {Triaud}, {Hellier},
  {Maxted}, {Pollaco}, {Queloz}, {Smalley}, {West}, {Wilson}, {Bentley},
  {Collier Cameron}, {Enoch}, {Hebb}, {Horne}, {Irwin}, {Joshi}, {Lister},
  {Mayor}, {Pepe}, {Parley}, {Segransan}, {Udry}, \& {Wheatley}}]{Gillion+2009}
{Gillon}, M., {Anderson}, D.~R., {Triaud}, A.~H.~M.~J., {et~al.} 2009, \aap,
  501, 785, \dodoi{10.1051/0004-6361/200911749}

\bibitem[{{H{\'e}brard} {et~al.}(2008){H{\'e}brard}, {Bouchy}, {Pont},
  {Loeillet}, {Rabus}, {Bonfils}, {Moutou}, {Boisse}, {Delfosse}, {Desort},
  {Eggenberger}, {Ehrenreich}, {Forveille}, {Lagrange}, {Lovis}, {Mayor},
  {Pepe}, {Perrier}, {Queloz}, {Santos}, {S{\'e}gransan}, {Udry}, \&
  {Vidal-Madjar}}]{hebrard2008}
{H{\'e}brard}, G., {Bouchy}, F., {Pont}, F., {et~al.} 2008, \aap, 488, 763,
  \dodoi{10.1051/0004-6361:200810056}

\bibitem[{{H{\'e}brard} {et~al.}(2011{\natexlab{a}}){H{\'e}brard},
  {Ehrenreich}, {Bouchy}, {Delfosse}, {Moutou}, {Arnold}, {Boisse}, {Bonfils},
  {D{\'{\i}}az}, {Eggenberger}, {Forveille}, {Lagrange}, {Lovis}, {Pepe},
  {Perrier}, {Queloz}, {Santerne}, {Santos}, {S{\'e}gransan}, {Udry}, \&
  {Vidal-Madjar}}]{hebrard2011b}
{H{\'e}brard}, G., {Ehrenreich}, D., {Bouchy}, F., {et~al.} 2011{\natexlab{a}},
  \aap, 527, L11, \dodoi{10.1051/0004-6361/201016331}

\bibitem[{{H{\'e}brard} {et~al.}(2011{\natexlab{b}}){H{\'e}brard}, {Evans},
  {Alonso}, {Fridlund}, {Ofir}, {Aigrain}, {Guillot}, {Almenara}, {Auvergne},
  {Baglin}, {Barge}, {Bonomo}, {Bord{\'e}}, {Bouchy}, {Cabrera}, {Carone},
  {Carpano}, {Cavarroc}, {Csizmadia}, {Deeg}, {Deleuil}, {D{\'{\i}}az},
  {Dvorak}, {Erikson}, {Ferraz-Mello}, {Gandolfi}, {Gibson}, {Gillon},
  {Guenther}, {Hatzes}, {Havel}, {Jorda}, {Lammer}, {L{\'e}ger}, {Llebaria},
  {Mazeh}, {Moutou}, {Ollivier}, {Parviainen}, {P{\"a}tzold}, {Queloz},
  {Rauer}, {Rouan}, {Santerne}, {Schneider}, {Tingley}, \&
  {Wuchterl}}]{Hebrard+2011}
{H{\'e}brard}, G., {Evans}, T.~M., {Alonso}, R., {et~al.} 2011{\natexlab{b}},
  \aap, 533, A130, \dodoi{10.1051/0004-6361/201117192}

\bibitem[{{Hellier} {et~al.}(2019){Hellier}, {Anderson}, {Triaud}, {Bouchy},
  {Burdanov}, {Collier Cameron}, {Delrez}, {Ehrenreich}, {Gillon}, {Jehin},
  {Lendl}, {Linder}, {Nielsen}, {Maxted}, {Pepe}, {Pollacco}, {Queloz},
  {S{\'e}gransan}, {Smalley}, {Spake}, {Temple}, {Udry}, {West}, \&
  {Wyttenbach}}]{Hellier+2019}
{Hellier}, C., {Anderson}, D.~R., {Triaud}, A.~H.~M.~J., {et~al.} 2019, \mnras,
  488, 3067, \dodoi{10.1093/mnras/stz1903}

\bibitem[{{Henry} \& {Winn}(2008)}]{HenryWinn2008}
{Henry}, G.~W., \& {Winn}, J.~N. 2008, \aj, 135, 68,
  \dodoi{10.1088/0004-6256/135/1/68}

\bibitem[{{Hirano} {et~al.}(2020{\natexlab{a}}){Hirano}, {Krishnamurthy},
  {Gaidos}, {Flewelling}, {Mann}, {Narita}, {Plavchan}, {Kotani}, {Tamura},
  {Harakawa}, {Hodapp}, {Ishizuka}, {Jacobson}, {Konishi}, {Kudo}, {Kurokawa},
  {Kuzuhara}, {Nishikawa}, {Omiya}, {Serizawa}, {Ueda}, \&
  {Vievard}}]{Hirano+2020}
{Hirano}, T., {Krishnamurthy}, V., {Gaidos}, E., {et~al.} 2020{\natexlab{a}},
  arXiv e-prints, arXiv:2006.13243.
\newblock \doarXiv{2006.13243}

\bibitem[{{Hirano} {et~al.}(2020{\natexlab{b}}){Hirano}, {Gaidos}, {Winn},
  {Dai}, {Fukui}, {Kuzuhara}, {Kotani}, {Tamura}, {Hjorth}, {Albrecht},
  {Huber}, {Bolmont}, {Harakawa}, {Hodapp}, {Ishizuka}, {Jacobson}, {Konishi},
  {Kudo}, {Kurokawa}, {Nishikawa}, {Omiya}, {Serizawa}, {Ueda}, \&
  {Weiss}}]{Hirano+2020b}
{Hirano}, T., {Gaidos}, E., {Winn}, J.~N., {et~al.} 2020{\natexlab{b}}, arXiv
  e-prints, arXiv:2002.05892.
\newblock \doarXiv{2002.05892}

\bibitem[{{Hjorth} {et~al.}(2021){Hjorth}, {Albrecht}, {Hirano}, {Winn},
  {Dawson}, {Zanazzi}, {Knudstrup}, \& {Sato}}]{Hjorth+2021}
{Hjorth}, M., {Albrecht}, S., {Hirano}, T., {et~al.} 2021, Proceedings of the
  National Academy of Science, 118, 2017418118, \dodoi{10.1073/pnas.2017418118}

\bibitem[{{Hjorth} {et~al.}(2019){Hjorth}, {Justesen}, {Hirano}, {Albrecht},
  {Gandolfi}, {Dai}, {Alonso}, {Barrag{\'a}n}, {Esposito}, {Kuzuhara}, {Lam},
  {Livingston}, {Montanes-Rodriguez}, {Narita}, {Nowak}, {Prieto-Arranz},
  {Redfield}, {Rodler}, {Van Eylen}, {Winn}, {Antoniciello}, {Cabrera},
  {Cochran}, {Csizmadia}, {de Leon}, {Deeg}, {Eigm{\"u}ller}, {Endl},
  {Erikson}, {Fridlund}, {Grziwa}, {Guenther}, {Hatzes}, {Heeren}, {Hidalgo},
  {Korth}, {Luque}, {Nespral}, {Palle}, {P{\"a}tzold}, {Persson}, {Rauer},
  {Smith}, \& {Trifonov}}]{Hjorth+2019}
{Hjorth}, M., {Justesen}, A.~B., {Hirano}, T., {et~al.} 2019, \mnras, 484,
  3522, \dodoi{10.1093/mnras/stz139}

\bibitem[{{Howarth} \& {Morello}(2017)}]{HowarthMorello2017}
{Howarth}, I.~D., \& {Morello}, G. 2017, \mnras, 470, 932,
  \dodoi{10.1093/mnras/stx1260}

\bibitem[{Hunter(2007)}]{Hunter:2007}
Hunter, J.~D. 2007, Computing in Science \& Engineering, 9, 90,
  \dodoi{10.1109/MCSE.2007.55}

\bibitem[{{Johnson} {et~al.}(2017){Johnson}, {Cochran}, {Addison}, {Tinney}, \&
  {Wright}}]{Johnson+2017}
{Johnson}, M.~C., {Cochran}, W.~D., {Addison}, B.~C., {Tinney}, C.~G., \&
  {Wright}, D.~J. 2017, \aj, 154, 137, \dodoi{10.3847/1538-3881/aa8462}

\bibitem[{{Johnson} {et~al.}(2015){Johnson}, {Cochran}, {Collier Cameron}, \&
  {Bayliss}}]{Johnson+2015}
{Johnson}, M.~C., {Cochran}, W.~D., {Collier Cameron}, A., \& {Bayliss}, D.
  2015, \apjl, 810, L23, \dodoi{10.1088/2041-8205/810/2/L23}

\bibitem[{{Kunovac Hod{\v{z}}i{\'c}} {et~al.}(2021){Kunovac Hod{\v{z}}i{\'c}},
  {Triaud}, {Cegla}, {Chaplin}, \& {Davies}}]{Kunovac_-Hodzic+2021}
{Kunovac Hod{\v{z}}i{\'c}}, V., {Triaud}, A. H.~M.~J., {Cegla}, H.~M.,
  {Chaplin}, W.~J., \& {Davies}, G.~R. 2021, \mnras, 502, 2893,
  \dodoi{10.1093/mnras/stab237}

\bibitem[{{Lai}(2012)}]{lai2012}
{Lai}, D. 2012, \mnras, 423, 486, \dodoi{10.1111/j.1365-2966.2012.20893.x}

\bibitem[{{Lai} {et~al.}(2011){Lai}, {Foucart}, \& {Lin}}]{lai2011}
{Lai}, D., {Foucart}, F., \& {Lin}, D.~N.~C. 2011, \mnras, 412, 2790,
  \dodoi{10.1111/j.1365-2966.2010.18127.x}

\bibitem[{{Lendl} {et~al.}(2020){Lendl}, {Csizmadia}, {Deline}, {Fossati},
  {Kitzmann}, {Heng}, {Hoyer}, {Salmon}, {Benz}, {Broeg}, {Ehrenreich},
  {Fortier}, {Queloz}, {Bonfanti}, {Brandeker}, {Collier Cameron}, {Delrez},
  {Garcia Mu{\~n}oz}, {Hooton}, {Maxted}, {Morris}, {Van Grootel}, {Wilson},
  {Alibert}, {Alonso}, {Asquier}, {Bandy}, {B{\'a}rczy}, {Barrado}, {Barros},
  {Baumjohann}, {Beck}, {Beck}, {Bekkelien}, {Bergomi}, {Billot}, {Biondi},
  {Bonfils}, {Bourrier}, {Busch}, {Cabrera}, {Cessa}, {Charnoz}, {Chazelas},
  {Corral Van Damme}, {Davies}, {Deleuil}, {Demangeon}, {Demory}, {Erikson},
  {Farinato}, {Fridlund}, {Futyan}, {Gandolfi}, {Gillon}, {Guterman}, {Hasiba},
  {Hernandez}, {Isaak}, {Kiss}, {Kuntzer}, {Lecavelier des Etangs},
  {L{\"u}ftinger}, {Laskar}, {Lovis}, {Magrin}, {Malvasio}, {Marafatto},
  {Michaelis}, {Munari}, {Nascimbeni}, {Olofsson}, {Ottacher}, {Ottensamer},
  {Pagano}, {Pall{\'e}}, {Peter}, {Piazza}, {Piotto}, {Pollacco}, {Ratti},
  {Rauer}, {Ragazzoni}, {Rando}, {Ribas}, {Rieder}, {Rohlfs}, {Safa}, {Santos},
  {Scandariato}, {S{\'e}gransan}, {Simon}, {Singh}, {Smith}, {Sordet}, {Sousa},
  {Steller}, {Szab{\'o}}, {Thomas}, {Tschentscher}, {Udry}, {Viotto}, {Walter},
  {Walton}, {Wildi}, \& {Wolter}}]{Lendl+2020}
{Lendl}, M., {Csizmadia}, S., {Deline}, A., {et~al.} 2020, \aap, 643, A94,
  \dodoi{10.1051/0004-6361/202038677}

\bibitem[{{Louden} {et~al.}(2021){Louden}, {Winn}, {Petigura}, {Isaacson},
  {Howard}, {Masuda}, {Albrecht}, \& {Kosiarek}}]{Louden+2021}
{Louden}, E.~M., {Winn}, J.~N., {Petigura}, E.~A., {et~al.} 2021, \aj, 161, 68,
  \dodoi{10.3847/1538-3881/abcebd}

\bibitem[{{Lund} {et~al.}(2014){Lund}, {Lundkvist}, {Silva Aguirre}, {Houdek},
  {Casagrande}, {Van Eylen}, {Campante}, {Karoff}, {Kjeldsen}, {Albrecht},
  {Chaplin}, {Nielsen}, {Degroote}, {Davies}, \& {Handberg}}]{Lund+2014}
{Lund}, M.~N., {Lundkvist}, M., {Silva Aguirre}, V., {et~al.} 2014, \aap, 570,
  A54, \dodoi{10.1051/0004-6361/201424326}

\bibitem[{{Mancini} {et~al.}(2015){Mancini}, {Esposito}, {Covino}, {Raia},
  {Southworth}, {Tregloan-Reed}, {Biazzo}, {Bonomo}, {Desidera}, {Lanza},
  {Maciejewski}, {Poretti}, {Sozzetti}, {Borsa}, {Bruni}, {Ciceri}, {Claudi},
  {Cosentino}, {Gratton}, {Martinez Fiorenzano}, {Lodato}, {Lorenzi},
  {Marzari}, {Murabito}, {Affer}, {Bignamini}, {Bedin}, {Boccato}, {Damasso},
  {Henning}, {Maggio}, {Micela}, {Molinari}, {Pagano}, {Piotto}, {Rainer},
  {Scandariato}, {Smareglia}, \& {Zanmar Sanchez}}]{Mancini+2015b}
{Mancini}, L., {Esposito}, M., {Covino}, E., {et~al.} 2015, \aap, 579, A136,
  \dodoi{10.1051/0004-6361/201526030}

\bibitem[{{Mancini} {et~al.}(2018){Mancini}, {Esposito}, {Covino},
  {Southworth}, {Biazzo}, {Bruni}, {Ciceri}, {Evans}, {Lanza}, {Poretti},
  {Sarkis}, {Smith}, {Brogi}, {Affer}, {Benatti}, {Bignamini}, {Boccato},
  {Bonomo}, {Borsa}, {Carleo}, {Claudi}, {Cosentino}, {Damasso}, {Desidera},
  {Giacobbe}, {Gonzalez-Alvarez}, {Gratton}, {Harutyunyan}, {Leto}, {Maggio},
  {Malavolta}, {Maldonado}, {Martinez-Fiorenzano}, {Masiero}, {Micela},
  {Molinari}, {Nascimbeni}, {Pagano}, {Pedani}, {Piotto}, {Rainer},
  {Scandariato}, {Smareglia}, {Sozzetti}, {Andreuzzi}, \&
  {Henning}}]{Mancini+2018}
---. 2018, ArXiv e-prints.
\newblock \doarXiv{1802.03859}

\bibitem[{{Mann} {et~al.}(2020){Mann}, {Johnson}, {Vanderburg}, {Kraus},
  {Rizzuto}, {Wood}, {Bush}, {Rockcliffe}, {Newton}, {Latham}, {Mamajek},
  {Zhou}, {Quinn}, {Thao}, {Benatti}, {Cosentino}, {Desidera}, {Harutyunyan},
  {Lovis}, {Mortier}, {Pepe}, {Poretti}, {Wilson}, {Kristiansen}, {Gagliano},
  {Jacobs}, {LaCourse}, {Omohundro}, {Schwengeler}, {Kane}, {Hill}, {Rabus},
  {Esquerdo}, {Berlind}, {Collins}, {Murawski}, {Aitken}, {Hazam Sallam},
  {Massey}, {Ricker}, {Vanderspek}, {Seager}, {Winn}, {Jenkins}, {Barclay},
  {Caldwell}, {Dragomir}, {Doty}, {Glidden}, {Tenenbaum}, {Torres}, {Twicken},
  \& {Villanueva}}]{Mann+2020}
{Mann}, A.~W., {Johnson}, M.~C., {Vanderburg}, A., {et~al.} 2020, arXiv
  e-prints, arXiv:2005.00047.
\newblock \doarXiv{2005.00047}

\bibitem[{{Masuda}(2015)}]{Masuda2015}
{Masuda}, K. 2015, \apj, 805, 28, \dodoi{10.1088/0004-637X/805/1/28}

\bibitem[{{Masuda} \& {Winn}(2020)}]{MasudaWinn2020}
{Masuda}, K., \& {Winn}, J.~N. 2020, \aj, 159, 81,
  \dodoi{10.3847/1538-3881/ab65be}

\bibitem[{{Maxted} {et~al.}(2015){Maxted}, {Serenelli}, \&
  {Southworth}}]{MaxtedSerenelliSouthworth2015}
{Maxted}, P.~F.~L., {Serenelli}, A.~M., \& {Southworth}, J. 2015, \aap, 577,
  A90, \dodoi{10.1051/0004-6361/201525774}

\bibitem[{{Mazeh} {et~al.}(2015){Mazeh}, {Perets}, {McQuillan}, \&
  {Goldstein}}]{mazeh2015}
{Mazeh}, T., {Perets}, H.~B., {McQuillan}, A., \& {Goldstein}, E.~S. 2015,
  \apj, 801, 3, \dodoi{10.1088/0004-637X/801/1/3}

\bibitem[{{McQuillan} {et~al.}(2013){McQuillan}, {Mazeh}, \&
  {Aigrain}}]{McQuillanMazehAigrain2013}
{McQuillan}, A., {Mazeh}, T., \& {Aigrain}, S. 2013, \apjl, 775, L11,
  \dodoi{10.1088/2041-8205/775/1/L11}

\bibitem[{{Mislis} {et~al.}(2015){Mislis}, {Mancini}, {Tregloan-Reed},
  {Ciceri}, {Southworth}, {D'Ago}, {Bruni}, {Ba{\c{s}}t{\"u}rk}, {Alsubai},
  {Bachelet}, {Bramich}, {Henning}, {Hinse}, {Iannella}, {Parley}, \&
  {Schroeder}}]{Mislis+2015}
{Mislis}, D., {Mancini}, L., {Tregloan-Reed}, J., {et~al.} 2015, \mnras, 448,
  2617, \dodoi{10.1093/mnras/stv197}

\bibitem[{{Mohler-Fischer} {et~al.}(2013){Mohler-Fischer}, {Mancini},
  {Hartman}, {Bakos}, {Penev}, {Bayliss}, {Jord{\'a}n}, {Csubry}, {Zhou},
  {Rabus}, {Nikolov}, {Brahm}, {Espinoza}, {Buchhave}, {B{\'e}ky}, {Suc},
  {Cs{\'a}k}, {Henning}, {Wright}, {Tinney}, {Addison}, {Schmidt}, {Noyes},
  {Papp}, {L{\'a}z{\'a}r}, {S{\'a}ri}, \& {Conroy}}]{Mohler-Fischer+2013}
{Mohler-Fischer}, M., {Mancini}, L., {Hartman}, J.~D., {et~al.} 2013, \aap,
  558, A55, \dodoi{10.1051/0004-6361/201321663}

\bibitem[{{Mo{\v{c}}nik} {et~al.}(2016){Mo{\v{c}}nik}, {Clark}, {Anderson},
  {Hellier}, \& {Brown}}]{Mocnik+2016}
{Mo{\v{c}}nik}, T., {Clark}, B.~J.~M., {Anderson}, D.~R., {Hellier}, C., \&
  {Brown}, D.~J.~A. 2016, \aj, 151, 150, \dodoi{10.3847/0004-6256/151/6/150}

\bibitem[{{Mo{\v{c}}nik} {et~al.}(2017){Mo{\v{c}}nik}, {Southworth}, \&
  {Hellier}}]{MocnikSouthworthHellier2017}
{Mo{\v{c}}nik}, T., {Southworth}, J., \& {Hellier}, C. 2017, \mnras, 471, 394,
  \dodoi{10.1093/mnras/stx1557}

\bibitem[{{Naoz} {et~al.}(2011){Naoz}, {Farr}, {Lithwick}, {Rasio}, \&
  {Teyssandier}}]{naoz2011}
{Naoz}, S., {Farr}, W.~M., {Lithwick}, Y., {Rasio}, F.~A., \& {Teyssandier}, J.
  2011, \nat, 473, 187, \dodoi{10.1038/nature10076}

\bibitem[{{Neveu-VanMalle} {et~al.}(2014){Neveu-VanMalle}, {Queloz},
  {Anderson}, {Charbonnel}, {Collier Cameron}, {Delrez}, {Gillon}, {Hellier},
  {Jehin}, {Lendl}, {Maxted}, {Pepe}, {Pollacco}, {S{\'e}gransan}, {Smalley},
  {Smith}, {Southworth}, {Triaud}, {Udry}, \& {West}}]{Neveu-VanMalle+2014}
{Neveu-VanMalle}, M., {Queloz}, D., {Anderson}, D.~R., {et~al.} 2014, \aap,
  572, A49, \dodoi{10.1051/0004-6361/201424744}

\bibitem[{{Perryman}(2011)}]{perryman2011}
{Perryman}, M. 2011, {The Exoplanet Handbook}, ed. {Perryman, M.}

\bibitem[{{Petrovich} {et~al.}(2020){Petrovich}, {Mu{\~n}oz}, {Kratter}, \&
  {Malhotra}}]{petrovich2020}
{Petrovich}, C., {Mu{\~n}oz}, D.~J., {Kratter}, K.~M., \& {Malhotra}, R. 2020,
  \apjl, 902, L5, \dodoi{10.3847/2041-8213/abb952}

\bibitem[{{Queloz} {et~al.}(2010){Queloz}, {Anderson}, {Collier Cameron},
  {Gillon}, {Hebb}, {Hellier}, {Maxted}, {Pepe}, {Pollacco}, {S{\'e}gransan},
  {Smalley}, {Triaud}, {Udry}, \& {West}}]{queloz2010}
{Queloz}, D., {Anderson}, D., {Collier Cameron}, A., {et~al.} 2010, \aap, 517,
  L1, \dodoi{10.1051/0004-6361/201014768}

\bibitem[{{Rogers} \& {Lin}(2013)}]{rogers2013b}
{Rogers}, T.~M., \& {Lin}, D.~N.~C. 2013, \apjl, 769, L10,
  \dodoi{10.1088/2041-8205/769/1/L10}

\bibitem[{{Romanova} {et~al.}(2020){Romanova}, {Koldoba}, {Ustyugova},
  {Blinova}, {Lai}, \& {Lovelace}}]{romanova2020}
{Romanova}, M.~M., {Koldoba}, A.~V., {Ustyugova}, G.~V., {et~al.} 2020, arXiv
  e-prints, arXiv:2012.10826.
\newblock \doarXiv{2012.10826}

\bibitem[{{Rosich} {et~al.}(2020){Rosich}, {Herrero}, {Mallonn}, {Ribas},
  {Morales}, {Perger}, {Anglada-Escud{\'e}}, \& {Granzer}}]{Rosich+2020}
{Rosich}, A., {Herrero}, E., {Mallonn}, M., {et~al.} 2020, \aap, 641, A82,
  \dodoi{10.1051/0004-6361/202037586}

\bibitem[{{Rubenzahl} {et~al.}(2021){Rubenzahl}, {Dai}, {Howard}, {Chontos},
  {Giacalone}, {Lubin}, {Rosenthal}, {Isaacson}, {Batalha}, {Crossfield},
  {Dressing}, {Fulton}, {Huber}, {Kane}, {Petigura}, {Robertson}, {Roy},
  {Weiss}, {Beard}, {Hill}, {Mayo}, {Mo{\v{c}}nik}, {Akana Murphy}, \&
  {Scarsdale}}]{Rubenzahl+2021}
{Rubenzahl}, R.~A., {Dai}, F., {Howard}, A.~W., {et~al.} 2021, arXiv e-prints,
  arXiv:2101.09371.
\newblock \doarXiv{2101.09371}

\bibitem[{{Sanchis-Ojeda} \& {Winn}(2011)}]{Sanchis-OjedaWinn2011}
{Sanchis-Ojeda}, R., \& {Winn}, J.~N. 2011, \apj, 743, 61,
  \dodoi{10.1088/0004-637X/743/1/61}

\bibitem[{{Sanchis-Ojeda} {et~al.}(2011){Sanchis-Ojeda}, {Winn}, {Holman},
  {Carter}, {Osip}, \& {Fuentes}}]{Sanchis-Ojeda+2011}
{Sanchis-Ojeda}, R., {Winn}, J.~N., {Holman}, M.~J., {et~al.} 2011, \apj, 733,
  127, \dodoi{10.1088/0004-637X/733/2/127}

\bibitem[{{Sanchis-Ojeda} {et~al.}(2013){Sanchis-Ojeda}, {Winn}, {Marcy},
  {Howard}, {Isaacson}, {Johnson}, {Torres}, {Albrecht}, {Campante}, {Chaplin},
  {Davies}, {Lund}, {Carter}, {Dawson}, {Buchhave}, {Everett}, {Fischer},
  {Geary}, {Gilliland}, {Horch}, {Howell}, \& {Latham}}]{Sanchis-Ojeda+2013}
{Sanchis-Ojeda}, R., {Winn}, J.~N., {Marcy}, G.~W., {et~al.} 2013, \apj, 775,
  54, \dodoi{10.1088/0004-637X/775/1/54}

\bibitem[{{Santerne} {et~al.}(2016){Santerne}, {H{\'e}brard}, {Lillo-Box},
  {Armstrong}, {Barros}, {Demangeon}, {Barrado}, {Debackere}, {Deleuil},
  {Delgado Mena}, {Montalto}, {Pollacco}, {Osborn}, {Sousa}, {Abe},
  {Adibekyan}, {Almenara}, {Andr{\'e}}, {Arlic}, {Barthe}, {Bendjoya},
  {Behrend}, {Boisse}, {Bouchy}, {Boussier}, {Bretton}, {Brown}, {Carry},
  {Cailleau}, {Conseil}, {Coulon}, {Courcol}, {Dauchet}, {Dalouzy}, {Deldem},
  {Desormi{\`e}res}, {Dubreuil}, {Fehrenbach}, {Ferratfiat}, {Girelli},
  {Gregorio}, {Jaecques}, {Kugel}, {Kirk}, {Labrevoir}, {Lachuri{\'e}}, {Lam},
  {Le Guen}, {Martinez}, {Maurin}, {McCormac}, {Pioppa}, {Quadri},
  {Rajpurohit}, {Rey}, {Rivet}, {Roy}, {Santos}, {Signoret}, {Strabla},
  {Suarez}, {Toublanc}, {Tsantaki}, {Vienney}, {Wilson}, {Bachschmidt},
  {Colas}, {Gerteis}, {Louis}, {Mario}, {Marlot}, {Montier}, {Perroud}, {Pic},
  {Romeuf}, {Ubaud}, \& {Verilhac}}]{Santerne+2016}
{Santerne}, A., {H{\'e}brard}, G., {Lillo-Box}, J., {et~al.} 2016, \apj, 824,
  55, \dodoi{10.3847/0004-637X/824/1/55}

\bibitem[{{Santos} {et~al.}(2020){Santos}, {Cristo}, {Demangeon}, {Oshagh},
  {Allart}, {Barros}, {Borsa}, {Bourrier}, {Casasayas-Barris}, {Ehrenreich},
  {Faria}, {Figueira}, {Martins}, {Micela}, {Pall{\'e}}, {Sozzetti},
  {Tabernero}, {Zapatero Osorio}, {Pepe}, {Cristiani}, {Rebolo}, {Adibekyan},
  {Allende Prieto}, {Alibert}, {Bouchy}, {Cabral}, {Dekker}, {Di Marcantonio},
  {D'Odorico}, {Dumusque}, {Gonz{\'a}lez Hern{\'a}ndez}, {Lavie}, {Lo Curto},
  {Lovis}, {Manescau}, {Martins}, {M{\'e}gevand}, {Mehner}, {Molaro}, {Nunes},
  {Poretti}, {Riva}, {Sousa}, {Su{\'a}rez Mascare{\~n}o}, \&
  {Udry}}]{Santos+2020}
{Santos}, N.~C., {Cristo}, E., {Demangeon}, O., {et~al.} 2020, \aap, 644, A51,
  \dodoi{10.1051/0004-6361/202039454}

\bibitem[{{Schlaufman}(2010)}]{schlaufman2010}
{Schlaufman}, K.~C. 2010, \apj, 719, 602, \dodoi{10.1088/0004-637X/719/1/602}

\bibitem[{{Shporer} {et~al.}(2014){Shporer}, {O'Rourke}, {Knutson},
  {Szab{\'o}}, {Zhao}, {Burrows}, {Fortney}, {Agol}, {Cowan}, {Desert},
  {Howard}, {Isaacson}, {Lewis}, {Showman}, \& {Todorov}}]{Shporer+2014}
{Shporer}, A., {O'Rourke}, J.~G., {Knutson}, H.~A., {et~al.} 2014, \apj, 788,
  92, \dodoi{10.1088/0004-637X/788/1/92}

\bibitem[{{Southworth} {et~al.}(2011){Southworth}, {Dominik}, {J{\o}rgensen},
  {Rahvar}, {Snodgrass}, {Alsubai}, {Bozza}, {Browne}, {Burgdorf}, {Calchi
  Novati}, {Dodds}, {Dreizler}, {Finet}, {Gerner}, {Hardis}, {Harps{\o}e},
  {Hellier}, {Hinse}, {Hundertmark}, {Kains}, {Kerins}, {Liebig}, {Mancini},
  {Mathiasen}, {Penny}, {Proft}, {Ricci}, {Sahu}, {Scarpetta}, {Sch{\"a}fer},
  {Sch{\"o}nebeck}, \& {Surdej}}]{southworth2011}
{Southworth}, J., {Dominik}, M., {J{\o}rgensen}, U.~G., {et~al.} 2011, \aap,
  527, A8, \dodoi{10.1051/0004-6361/201016183}

\bibitem[{{Southworth} {et~al.}(2016){Southworth}, {Tregloan-Reed}, {Andersen},
  {Calchi Novati}, {Ciceri}, {Colque}, {D'Ago}, {Dominik}, {Evans}, {Gu},
  {Herrera-Cordova}, {Hinse}, {J{\o}rgensen}, {Juncher}, {Kuffmeier},
  {Mancini}, {Peixinho}, {Popovas}, {Rabus}, {Skottfelt}, {Tronsgaard},
  {Unda-Sanzana}, {Wang}, {Wertz}, {Alsubai}, {Andersen}, {Bozza}, {Bramich},
  {Burgdorf}, {Damerdji}, {Diehl}, {Elyiv}, {Figuera Jaimes}, {Haugb{\o}lle},
  {Hundertmark}, {Kains}, {Kerins}, {Korhonen}, {Liebig}, {Mathiasen}, {Penny},
  {Rahvar}, {Scarpetta}, {Schmidt}, {Snodgrass}, {Starkey}, {Surdej}, {Vilela},
  {von Essen}, \& {Wang}}]{Southworth+2016}
{Southworth}, J., {Tregloan-Reed}, J., {Andersen}, M.~I., {et~al.} 2016,
  \mnras, 457, 4205, \dodoi{10.1093/mnras/stw279}

\bibitem[{{Stefansson} {et~al.}(2020){Stefansson}, {Mahadevan}, {Maney},
  {Ninan}, {Robertson}, {Rajagopal}, {Haase}, {Allen}, {Ford}, {Winn},
  {Wolfgang}, {Dawson}, {Wisniewski}, {Bender}, {Ca{\~n}as}, {Cochran},
  {Diddams}, {Fredrick}, {Halverson}, {Hearty}, {Hebb}, {Kanodia}, {Levi},
  {Metcalf}, {Monson}, {Ramsey}, {Roy}, {Schwab}, {Terrien}, \&
  {Wright}}]{Stefansson+2020}
{Stefansson}, G., {Mahadevan}, S., {Maney}, M., {et~al.} 2020, \aj, 160, 192,
  \dodoi{10.3847/1538-3881/abb13a}

\bibitem[{{Temple} {et~al.}(2017){Temple}, {Hellier}, {Albrow}, {Anderson},
  {Bayliss}, {Beatty}, {Bieryla}, {Brown}, {Cargile}, {Collier Cameron},
  {Collins}, {Col{\'o}n}, {Curtis}, {D'Ago}, {Delrez}, {Eastman}, {Gaudi},
  {Gillon}, {Gregorio}, {James}, {Jehin}, {Joner}, {Kielkopf}, {Kuhn},
  {Labadie-Bartz}, {Latham}, {Lendl}, {Lund}, {Malpas}, {Maxted}, {Myers},
  {Oberst}, {Pepe}, {Pepper}, {Pollacco}, {Queloz}, {Rodriguez},
  {S{\'e}gransan}, {Siverd}, {Smalley}, {Stassun}, {Stevens}, {Stockdale},
  {Tan}, {Triaud}, {Udry}, {Villanueva}, {West}, \& {Zhou}}]{Temple+2017}
{Temple}, L.~Y., {Hellier}, C., {Albrow}, M.~D., {et~al.} 2017, \mnras, 471,
  2743, \dodoi{10.1093/mnras/stx1729}

\bibitem[{{Tregloan-Reed} {et~al.}(2013){Tregloan-Reed}, {Southworth}, \&
  {Tappert}}]{Tregloan-reedSouthworthTappert2012}
{Tregloan-Reed}, J., {Southworth}, J., \& {Tappert}, C. 2013, \mnras, 428,
  3671, \dodoi{10.1093/mnras/sts306}

\bibitem[{{Tregloan-Reed} {et~al.}(2015){Tregloan-Reed}, {Southworth},
  {Burgdorf}, {Novati}, {Dominik}, {Finet}, {J{\o}rgensen}, {Maier}, {Mancini},
  {Prof}, {Ricci}, {Snodgrass}, {Bozza}, {Browne}, {Dodds}, {Gerner},
  {Harps{\o}e}, {Hinse}, {Hundertmark}, {Kains}, {Kerins}, {Liebig}, {Penny},
  {Rahvar}, {Sahu}, {Scarpetta}, {Sch{\"a}fer}, {Sch{\"o}nebeck}, {Skottfelt},
  \& {Surdej}}]{Tregloan-Reed+2015}
{Tregloan-Reed}, J., {Southworth}, J., {Burgdorf}, M., {et~al.} 2015, \mnras,
  450, 1760, \dodoi{10.1093/mnras/stv730}

\bibitem[{{Triaud}(2017)}]{triaud2017}
{Triaud}, A.~H.~M.~J. 2017, {The Rossiter-McLaughlin Effect in Exoplanet
  Research}, 2, \dodoi{10.1007/978-3-319-30648-3_2-1}

\bibitem[{{Triaud} {et~al.}(2010){Triaud}, {Collier Cameron}, {Queloz},
  {Anderson}, {Gillon}, {Hebb}, {Hellier}, {Loeillet}, {Maxted}, {Mayor},
  {Pepe}, {Pollacco}, {S{\'e}gransan}, {Smalley}, {Udry}, {West}, \&
  {Wheatley}}]{Triaud+2010}
{Triaud}, A.~H.~M.~J., {Collier Cameron}, A., {Queloz}, D., {et~al.} 2010,
  \aap, 524, A25, \dodoi{10.1051/0004-6361/201014525}

\bibitem[{{Vick} {et~al.}(2019){Vick}, {Lai}, \&
  {Anderson}}]{vick_lai_anderson2019}
{Vick}, M., {Lai}, D., \& {Anderson}, K.~R. 2019, \mnras,
  \dodoi{10.1093/mnras/stz354}

\bibitem[{{Wang} {et~al.}(2018){Wang}, {Addison}, {Fischer}, {Brewer},
  {Isaacson}, {Howard}, \& {Laughlin}}]{Wang+2018}
{Wang}, S., {Addison}, B., {Fischer}, D.~A., {et~al.} 2018, \aj, 155, 70,
  \dodoi{10.3847/1538-3881/aaa2fb}

\bibitem[{{Winn} {et~al.}(2009){Winn}, {Johnson}, {Fabrycky}, {Howard},
  {Marcy}, {Narita}, {Crossfield}, {Suto}, {Turner}, {Esquerdo}, \&
  {Holman}}]{Winn+2009_X03}
{Winn}, J.~N., {Johnson}, J.~A., {Fabrycky}, D., {et~al.} 2009, \apj, 700, 302,
  \dodoi{10.1088/0004-637X/700/1/302}

\bibitem[{{Winn} {et~al.}(2010){Winn}, {Johnson}, {Howard}, {Marcy},
  {Isaacson}, {Shporer}, {Bakos}, {Hartman}, \& {Albrecht}}]{Winn+2010_hatp11}
{Winn}, J.~N., {Johnson}, J.~A., {Howard}, A.~W., {et~al.} 2010, \apjl, 723,
  L223, \dodoi{10.1088/2041-8205/723/2/L223}

\bibitem[{{Wyttenbach} {et~al.}(2020){Wyttenbach}, {Molli{\`e}re},
  {Ehrenreich}, {Cegla}, {Bourrier}, {Lovis}, {Pino}, {Allart}, {Seidel},
  {Hoeijmakers}, {Nielsen}, {Lavie}, {Pepe}, {Bonfils}, \&
  {Snellen}}]{Wyttenbach+2020}
{Wyttenbach}, A., {Molli{\`e}re}, P., {Ehrenreich}, D., {et~al.} 2020, \aap,
  638, A87, \dodoi{10.1051/0004-6361/201937316}

\bibitem[{{Yu} {et~al.}(2018){Yu}, {Zhou}, {Rodriguez}, {Huang}, {Vanderburg},
  {Quinn}, {Gaudi}, {Beichman}, {Berlind}, {Bieryla}, {Calkins}, {Ciardi},
  {Crossfield}, {Eastman}, {Esquerdo}, {Latham}, {Stassun}, \&
  {Villanueva}}]{Yu+2018}
{Yu}, L., {Zhou}, G., {Rodriguez}, J.~E., {et~al.} 2018, \aj, 156, 250,
  \dodoi{10.3847/1538-3881/aae5d5}

\bibitem[{{Zhou} {et~al.}(2016){Zhou}, {Rodriguez}, {Collins}, {Beatty},
  {Oberst}, {Heintz}, {Stassun}, {Latham}, {Kuhn}, {Bieryla}, {Lund},
  {Labadie-Bartz}, {Siverd}, {Stevens}, {Gaudi}, {Pepper}, {Buchhave},
  {Eastman}, {Col{\'o}n}, {Cargile}, {James}, {Gregorio}, {Reed}, {Jensen},
  {Cohen}, {McLeod}, {Tan}, {Zambelli}, {Bayliss}, {Bento}, {Esquerdo},
  {Berlind}, {Calkins}, {Blancato}, {Manner}, {Samulski}, {Stockdale},
  {Nelson}, {Stephens}, {Curtis}, {Kielkopf}, {Fulton}, {DePoy}, {Marshall},
  {Pogge}, {Gould}, {Trueblood}, \& {Trueblood}}]{Zhou+2016b}
{Zhou}, G., {Rodriguez}, J.~E., {Collins}, K.~A., {et~al.} 2016, \aj, 152, 136,
  \dodoi{10.3847/0004-6256/152/5/136}

\bibitem[{{Zhou} {et~al.}(2020){Zhou}, {Winn}, {Newton}, {Quinn}, {Rodriguez},
  {Mann}, {Rizzuto}, {Vand erburg}, {Huang}, {Latham}, {Teske}, {Wang},
  {Shectman}, {Butler}, {Crane}, {Thompson}, {Henry}, {Paredes}, {Jao},
  {James}, \& {Hinojosa}}]{Zhou+2020}
{Zhou}, G., {Winn}, J.~N., {Newton}, E.~R., {et~al.} 2020, \apjl, 892, L21,
  \dodoi{10.3847/2041-8213/ab7d3c}

\bibitem[{{Zurlo} {et~al.}(2018){Zurlo}, {Mesa}, {Desidera}, {Messina},
  {Gratton}, {Moutou}, {Beuzit}, {Biller}, {Boccaletti}, {Bonavita},
  {Bonnefoy}, {Bhowmik}, {Brandner}, {Buenzli}, {Chauvin}, {Cudel}, {D'Orazi},
  {Feldt}, {Hagelberg}, {Janson}, {Lagrange}, {Langlois}, {Lannier}, {Lavie},
  {Lazzoni}, {Maire}, {Meyer}, {Mouillet}, {Peretti}, {Perrot}, {Potiron},
  {Salter}, {Schmidt}, {Sissa}, {Vigan}, {Delboulb{\'e}}, {Petit}, {Ramos},
  {Rigal}, \& {Rochat}}]{Zurlo+2018}
{Zurlo}, A., {Mesa}, D., {Desidera}, S., {et~al.} 2018, \mnras, 480, 35,
  \dodoi{10.1093/mnras/sty1809}

\end{thebibliography}
\end{document}